\documentclass{jfm_arxiv}
\usepackage{changes}
\usepackage{graphicx}
\usepackage{newtxtext}
\usepackage{newtxmath}
\usepackage{natbib}
\usepackage{hyperref}
\usepackage{color}
\usepackage{nicematrix}
\hypersetup{
    colorlinks = true,
    urlcolor   = blue,
    citecolor  = black,
}

\newcommand{\RomanNumeralCaps}[1]
\linenumbers

\def\d{\mathrm{d}}
\def\D{\mathrm{D}}
\def\bm{\boldsymbol}
\newcommand{\ed}[1]{\textcolor{red}{#1}}

\title{On buoyancy in disperse two-phase flow and its impact on the well-posedness of two-fluid models}

\author{Rui Zhu,\aff{1}
  Yulan Chen,\aff{1}
	Katharina Tholen,\aff{2}
  Zhiguo He,\aff{1}
	\corresp{\email{hezhiguo@zju.edu.cn}}
	\and Thomas P\"ahtz\aff{3}
	\corresp{\email{tpaehtz@gmail.com}}}

\affiliation{\aff{1}Institute of Port, Coastal and Offshore Engineering, Ocean College, Zhejiang University, 316021 Zhoushan, China
\aff{2}Institute for Theoretical Physics, Leipzig University, Br\"uderstra{\ss}e 16, 04103 Leipzig, Germany
\aff{3}School of Marine Science and Engineering, South China University of Technology, Guangzhou 511442, China}

\begin{document}
\maketitle

\begin{abstract}
The Maxey--Riley--Gatignol equation for the flow around a sphere at low particle Reynolds number tells us that the fluid-particle interaction force decomposes into a contribution from the local flow disturbance caused by the particle's boundary---consisting of the drag, virtual-mass, and history forces, and their Fax\'en corrections---and another contribution from the stress of the background flow, termed generalized-buoyancy force. There is also a consensus that, for general disperse two-phase flow, the interfacial force density, coupling the average fluid phase and dispersed-phase momentum balances, decomposes in a like manner. However, there has been a long-standing controversy about the physical closure separating the generalized-buoyancy from the interfacial force density, especially whether or not pseudo-stresses, such as the Reynolds stress, should be attributed to the background flow. Furthermore, most existing propositions for this closure involve small-particle approximations. Here, we show that all existing buoyancy closures are inconsistent with particle-resolving numerical simulations and/or at least one of two simple thought experiments designed to determine the roles of pseudo-stresses and small-particle approximations. We then derive the unique consistent closure. It requires no approximation and implies that all stresses and pseudo-stresses in the average fluid phase momentum balance, except the Reynolds stress, fully contribute to the background flow responsible for buoyancy. Remarkably, it exhibits a low-pass filter property, attenuating buoyancy at short wavelengths, that prevents it from causing Hadamard instabilities, constituting a first-principles-based solution to the long-standing ill-posedness problem of two-fluid models. When employing the derived closure, even simplistic two-fluid models are linearly well-posed.
\end{abstract}

\begin{keywords}
\end{keywords}

\section{Introduction} \label{Introduction}
Archimedes' principle states that a particle immersed in a stationary fluid experiences a force equal to the negative weight of the fluid it displaces~\citep{Batchelor00}: $\bm{F}_\mathrm{B}=-\int_{\mathbb{V}}\rho_\mathrm{f}\bm{g}\d V$, where $\rho_\mathrm{f}$ is the fluid density, $\bm{g}$ the gravitational acceleration, and $\mathbb{V}$ the submerged domain occupied by the particle. This study addresses the question of how to generalize this so-called buoyancy force to general dynamic fluid-particle systems, where the terms ``fluid'' and ``particle'' stand for, respectively, a phase of continuous fluid and a dispersed phase consisting of droplets, bubbles, and/or granular particles. Such systems are often modelled by effective two-fluid continuum models, in which buoyancy appears in the form of a generalized-buoyancy force density in the momentum balance equations \citep[e.g.,][]{Drew83,StewartWendroff84,Jackson00,Croweetal12,Lhuillieretal13,RevilBaudardChauchat13}. In spite of their widespread use, the partial differential equations (PDEs) underlying such models tend to be ill-posed due to unbounded growth of small-amplitude wave perturbations in the limit of vanishing wavelength \citep{StewartWendroff84,Lhuillieretal13,Panickeretal18,Langhametal25a}---a so-called Hadamard instability \citep{JosephSaut90}---and this buoyancy term has been recognized as the main culprit behind this unphysical behaviour \citep{Lhuillieretal13}. Most of the various solutions proposed to restore linear well-posedness of the PDEs, characterized by the absence of Hadamard instabilities, have consisted in introducing into the two-fluid models buoyancy-unrelated closure relations that dissipate short-wavelength growth contributions \citep{Lhuillieretal13,Breschetal18,Panickeretal18,Zhang21a,Fox19,Fox25}. However, to the best of our knowledge, no prior study has ever questioned whether the modelling of buoyancy itself may have been the main issue. In the present study, we arrive at that conclusion not by design but as a consequence of a first-principles-based analysis carried out to determine the expression that governs buoyancy in disperse two-phase flow. Our reasoning is built on a careful definition of the term ``buoyancy'' in dynamic systems, which we introduce below in a manner consistent with known results and common modelling practices.

To obtain a sensible definition of a generalized-buoyancy force, we start with the so-called Maxey--Riley--Gatignol (MRG) equation \citep{MaxeyRiley83,Gatignol83,Jaganathanetal25}, which was actually derived earlier by \citet{MazurBedeaux74}:
\begin{equation}
\begin{split}
 \bm{F}_\mathrm{h}=\underbrace{\int_\mathbb{V}\bm{\nabla}\cdot\bm{\tilde\sigma}\d V}_\text{force from background flow}&+\underbrace{\frac{3\eta_\mathrm{f}}{2R}\int_\mathbb{S}(\bm{\tilde u}-\bm{v}_\uparrow)\d S}_{\text{drag force}\;+\;\text{Fax\'en correction}}+\underbrace{\frac{\rho_\mathrm{f}}{2}\frac{\d}{\d t}\int_\mathbb{V}(\bm{\tilde u}-\bm{v}_\uparrow)\d V}_{\text{virtual-mass force}\;+\;\text{Fax\'en correction}} \\
 &+\underbrace{\frac{3\sqrt{\rho_\mathrm{f}\eta_\mathrm{f}}}{2\sqrt{\pi}}\int_{-\infty}^t\left[\frac{\frac{\d}{\d t^\prime}\int_\mathbb{S}(\bm{\tilde u}-\bm{v}_\uparrow)\d S}{\sqrt{t-t^\prime}}\right]\d t^\prime}_{\text{Basset history force}\;+\;\text{Fax\'en correction}}+\underbrace{0}_\text{lift force}.
\end{split} \label{MRG}
\end{equation}
It describes the time ($t$)-dependent hydrodynamic force $\bm{F}_\mathrm{h}$ on a single rigid sphere of radius $R$ moving at translational velocity $\bm{v}_\uparrow$ and submerged in a flow of an incompressible Newtonian fluid of density $\rho_\mathrm{f}$, dynamic viscosity $\eta_\mathrm{f}$, velocity $\bm{\tilde u}$, stress tensor $\bm{\tilde\sigma}=-\tilde P\bm{1}+\bm{\tilde\tau}$, pressure $\tilde P$, and shear stress tensor $\bm{\tilde\tau}=\eta_\mathrm{f}(\bm{\nabla}\bm{\tilde u}+(\bm{\nabla}\bm{\tilde u})^\mathrm{T})$. Here $\bm{1}$ is the identity tensor and a tilde indicates a quantity's value associated with the background flow, that is, the value it would have if the sphere were absent, i.e., if the no-slip boundary conditions at the sphere's surface $\mathbb{S}$ did not disturb the flow and if the sphere's interior $\mathbb{V}$ were replaced by fluid. Equation~(\ref{MRG}) represents the analytical solution for $\bm{F}_\mathrm{h}$ under the assumptions that the flow occupies an infinite, unbounded domain and that the particle Reynolds numbers $\Re_U\equiv\rho_\mathrm{f}UR^2/(L\eta_\mathrm{f})$ and $\Re_W\equiv\rho_\mathrm{f}WR/\eta_\mathrm{f}$ are vanishingly small \citep{MaxeyRiley83}, denoted as $\Re_U\sim0$ and $\Re_W\sim0$. Here $U$ and $W$ are characteristic scales for $|\bm{\tilde u}|$ and $|\bm{\tilde u}-\bm{v}_\uparrow|$, respectively, while $L$ is a characteristic scale for $|\bm{\tilde u}|/|\bm{\nabla}\bm{\tilde u}|$, i.e., the characteristic distance over which $|\bm{\tilde u}|$ changes. In (\ref{MRG}), $\bm{F}_\mathrm{h}$ is decomposed into the force from the undisturbed or background flow and the force from the disturbance flow, composed of the drag, virtual-mass, and history forces, and their Fax\'en corrections accounting for the inhomogeneity of $\bm{\tilde u}$ within $\mathbb{S}$ and $\mathbb{V}$; due to $\Re_U\sim0$ and $\Re_W\sim0$, the lift force vanishes relative to the other disturbance-flow forces \citep{MaxeyRiley83}. Note that the conditions $\Re_U\sim0$ and $\Re_W\sim0$ do not necessarily preclude the possibility that the background flow is turbulent \citep{MaxeyRiley83}.

In contrast to the above scenario, general disperse two-phase flow usually does not admit a rigorous solution. For this reason, one usually proposes semi-empirical closure relations for the average generalized-buoyancy, drag, virtual-mass, and lift forces \citep[e.g.,][]{DiFelice95,Jackson00,LothDorgan09,Croweetal12,ShiRzehak19,Jamshidietal21}. However, for consistency, it is required that a proposed set of closures is consistent with the MRG equation (\ref{MRG}) when $\Re_U\sim0$ and $\Re_W\sim0$. Therefore, since the drag, virtual-mass, and lift forces already have their respective counterparts in (\ref{MRG}), it makes sense to interpret the background-flow force in (\ref{MRG}) as the generalized-buoyancy force:
\begin{equation}
 \bm{F}_\mathrm{B}\equiv\int_{\mathbb{V}}\bm{\nabla}\cdot\bm{\tilde\sigma}\d V. \label{Buoyancy1}
\end{equation}
The alternative, common pressure-gradient-based definition $\bm{F}_\mathrm{B}=-\int_{\mathbb{V}}\bm{\nabla}\tilde P\d V$ \citep[e.g.,][]{Drew83,DiFelice95,Croweetal12,Lhuillieretal13,Fox19} is inappropriate, since it leaves the force $\int_{\mathbb{V}}\bm{\nabla}\cdot\bm{\tilde\tau}\d V$ as a remainder that is usually not being addressed in two-phase flow closures even though it clearly exists. This shear stress gradient force typically acts in the flow direction, i.e., it is not the ``missing'' lift force in (\ref{MRG}). Note that, in concert with (\ref{Buoyancy1}), we term the total force due to the disturbance flow, $\bm{F}_\mathrm{D}\equiv\bm{F}_\mathrm{h}-\bm{F}_\mathrm{B}$, the ``generalized-drag force'' for the scenario described by (\ref{MRG}).

Using the background flow momentum balance, (\ref{Buoyancy1}) is equivalent to
\begin{equation}
 \bm{F}_\mathrm{B}=\int_{\mathbb{V}}\rho_\mathrm{f}(\tilde{\D}_t\bm{\tilde u}-\bm{g})\d V, \label{Buoyancy2}
\end{equation}
where $\tilde\D_t\equiv\partial_t+\bm{\tilde u}\cdot\bm{\nabla}$ is the material derivative associated with the background flow. Equation~(\ref{Buoyancy2}) constitutes a modification of the buoyancy force expression for the stationary case by the term involving $\tilde{\D}_t\bm{\tilde u}$, which accounts for fluid inertia and ensures that $\bm{F}_\mathrm{B}$ is independent of coordinate transformations into accelerated frames of reference. In particular, for a steady, uniform Stokes flow down an inclined plane, (\ref{Buoyancy2}) yields $\bm{F}_\mathrm{B}=-\int_{\mathbb{V}}\rho_\mathrm{f}\bm{g}\d V$, whereas a pressure-gradient-based definition would have resulted in $\bm{F}_\mathrm{B}=-\int_{\mathbb{V}}\rho_\mathrm{f}g_z\bm{e_z}\d V$, different from the stationary case, where $\bm{e_z}$ is the unit vector in the direction ($z$) normal to the plane.

Now, a question that has sparked considerable debate over the past six decades is: What is the background flow stress responsible for buoyancy in general disperse two-phase flow? In other words, what is the analogue to $\bm{\tilde\sigma}$ that governs the generalized-buoyancy force on an individual particle of a general fluid-particle system through an equation analogous to (\ref{Buoyancy1})? The answer to this question is not trivial, since, when the particle Reynolds numbers $\Re_U$ and $\Re_W$ are large, nonlinear terms involving the background flow appear in the disturbance flow balance equations, meaning that the background and disturbance flows are no longer decoupled \citep{MaxeyRiley83}. In addition, closure relations for the generalized-drag force in turbulent particle-laden flows are usually expressed in terms of flow properties that have been averaged in some manner, such as an averaged flow velocity, or whose very existence is tied to some form of averaging, such as the fluid ($\beta_\mathrm{f}$) and dispersed-phase ($\beta_\mathrm{s}=1-\beta_\mathrm{f}$) volume fractions \citep{DiFelice95,Jackson00}. As a consequence, the background flow stress responsible for buoyancy should be somehow related to the averaged flow. However, averaging gives rise to two kinds of pseudo-stresses in the resulting average fluid phase momentum balance: the Reynolds stress $\bm{\sigma^\mathrm{f}_\mathrm{Re}}$ and a stress due to fluid-particle interactions, $\bm{\sigma^\mathrm{f}_\mathrm{s}}$, which is responsible for an effective viscosity increase and other effects \citep{Jackson97,Jackson00}. This raises the question: Which of these pseudo-stresses should be attributed to the background flow when modelling the generalized-buoyancy force? \citet{AndersonJackson67} and \citet{Jackson00} proposed that all pseudo-stresses fully contribute to the background flow stress responsible for buoyancy, whereas others considered only the fluid-phase-averaged stress \citep{Zhangetal07b,Maurinetal18}. In particular, \citet{Maurinetal18} argued strongly against including $\bm{\sigma^\mathrm{f}_\mathrm{Re}}$. Compromises between these two extremes have also been suggested. For example, \citet{RevilBaudardChauchat13} fully accounted for $\bm{\sigma^\mathrm{f}_\mathrm{s}}$ but disregarded $\bm{\sigma^\mathrm{f}_\mathrm{Re}}$. 

A key problem that has hindered conclusively answering the above question is that it is difficult to empirically test generalized-buoyancy expressions in isolation, since one usually has only access to the total hydrodynamic force or force density. In addition, Reynolds normal stresses tend to be almost negligible relative to the fluid-phase-averaged pressure in unidirectional systems, posing an obstacle to evaluating the effects of $\bm{\sigma^\mathrm{f}_\mathrm{Re}}$ in the gravity-aligned direction where buoyancy effects are usually most important. \citet{Lambetal17} are perhaps the only ones who attempted to circumvent these problems. They experimentally studied the average total hydrodynamic force acting on objects partially submerged in a shallow turbulent stream down an incline. Their data could be described by a standard drag force closure only if they assumed that the average generalized-buoyancy force $\langle\bm{F}_\mathrm{B}\rangle$, with $\langle\cdot\rangle$ the ensemble average, is aligned with the gravity direction, suggesting that $\langle\bm{F}_\mathrm{B}\rangle$ has a streamwise component. If they instead assumed that $\langle\bm{F}_\mathrm{B}\rangle$ were aligned with the direction perpendicular to the sediment bed, the resulting drag coefficients would have had slope dependencies these authors deemed unphysical (M.~P.~Lamb, personal communication 2024). Therefore, the measurements by \citet{Lambetal17} seem to support the notion that the Reynolds shear stress contributes to the background flow shear stress responsible for buoyancy.

Knowing the background flow stress responsible for buoyancy in general disperse two-phase flow is not the end of the story, since integral expressions such as (\ref{Buoyancy1}) can be of limited use and are often approximated as
\begin{equation}
 \int_{\mathbb{V}}\bm{\nabla}\cdot\bm{\sigma_\ast}\d V\approx V\bm{\nabla}\cdot\bm{\sigma_\ast}[{\bm{x}_\mathrm{o}}], \label{BuoyancySmall}
\end{equation}
where $V$ is the particle's volume, $\bm{x}_\mathrm{o}$ its centre of mass, and $\bm{\sigma_\ast}$ stands as a placeholder for any of the potential background flow stress tensors. Such kind of approximations, which assume that the particle size is sufficiently small, permit a simple translation of the average generalized-buoyancy force $\langle\bm{F}_\mathrm{B}\rangle$ acting on a particle into the corresponding interfacial force density $\beta_\mathrm{s}\langle\bm{f}_\mathrm{B}\rangle^\mathrm{s}$ appearing in two-fluid momentum balance equations: $\beta_\mathrm{s}\langle\bm{f}_\mathrm{B}\rangle^\mathrm{s}\approx\beta_\mathrm{s}\bm{\nabla}\cdot\bm{\sigma_\ast}$. However, to the best of our knowledge, the effects of such approximations have not been seriously studied in the literature. Note that, throughout this paper, we use upper-case letters ``$F$'' to indicate forces and lower-case letters ``$f$'' to indicate forces per unit volume.

It is also worth noting that some attempts to generalize buoyancy \citep[e.g.,][]{Cliftetal87} were based on expressions analogous to (\ref{Buoyancy2}) rather than (\ref{Buoyancy1}). While the two approaches are equivalent when considering an isolated particle in a pure-fluid flow (i.e., $\beta_\mathrm{s}\sim0$), they can result in very different outcomes for actual fluid-particle mixture flows, such as homogeneous air-fluidized beds where the effective fluid pressure gradient associated with the fluid-phase-averaged flow supports both the gravitational and generalized-drag force densities. In this case, an expression analogous to (\ref{Buoyancy2}) yields $\langle\bm{F}_\mathrm{B}\rangle=-\int_{\mathbb{V}}\rho_\mathrm{f}\bm{g}\d V$, whereas an expression analogous to (\ref{Buoyancy1}) results in $\langle\bm{F}_\mathrm{B}\rangle=-\int_{\mathbb{V}}\rho_\mathrm{m}\bm{g}\d V$ \citep{DiFelice95}, where $\rho_\mathrm{m}=\beta_\mathrm{f}\rho_\mathrm{f}+\beta_\mathrm{s}\rho_\mathrm{s}\gg\rho_\mathrm{f}$ is the mixture density, with $\rho_\mathrm{s}$ the dispersed-phase density. This sparked an intense debate in the 1980s, which was described in some detail by \citet{DiFelice95}. This debate was essentially resolved in favour of expressions involving $\rho_\mathrm{m}$, and therefore in favour of stress-based expressions analogous to (\ref{Buoyancy1}): It was shown that the hydrodynamic force acting on a coarse particle submerged in an air-fluidized bed of much finer particles can be expressed using a standard but $\rho_\mathrm{m}$-based closure for the drag force combined with $\langle\bm{F}_\mathrm{B}\rangle=-\int_{\mathbb{V}}\rho_\mathrm{m}\bm{g}\d V$, whereas $\rho_\mathrm{f}$-based approaches fail \citep{Rotondietal15}.

Here, we employ rigorous mathematical analysis, physical intuition, particle-resolving numerical simulations, and thought experiments to demonstrate inconsistencies in existing buoyancy closures and derive a new consistent closure. In \S\ref{TwoFluidMom}, we start with presenting the macroscopic two-fluid momentum balances and mathematically exact micromechanical expressions for the terms appearing therein, previously derived from averaging the fluid's and particles' equations of motion without application of approximations \citep{Pahtzetal26}. This section then defines the generalized-buoyancy force density and shows that a buoyancy closure based on the small-particle approximation (\ref{BuoyancySmall}) leads to an inconsistency for some fluid-particle systems at complete rest. It also lays out the physical and mathematical rationales behind previously proposed closures, and why we believe they are (or are not) appropriate. It ends with a physically inspired criterion that we propose a buoyancy closure should satisfy. Among other things, it states that the pseudo-stress $\bm{\sigma^\mathrm{f}_\mathrm{s}}$ fully contributes to the background flow stress responsible for buoyancy, whereas $\bm{\sigma^\mathrm{f}_\mathrm{Re}}$ should be disregarded, precisely as proposed by \citet{RevilBaudardChauchat13}. In \S\ref{PseudoStresses}, we present particle-resolving simulations and a thought experiment that strongly support this hypothesis, while disagreeing with others. In \S\ref{NewClosure}, by deriving the unique buoyancy closure that satisfies the criterion proposed in \S\ref{TwoFluidMom}, we generalize the closure by \citet{RevilBaudardChauchat13} to arbitrarily large particles. The crucial mathematical property of the resulting generalized closure is the attenuation of generalized-buoyancy force density contributions at short wavelengths. We show that this property prevents Hadamard instabilities even for simplistic two-fluid models (\S\ref{TwoFluidModels}).

\section{Two-fluid momentum balances and buoyancy closures} \label{TwoFluidMom}
\subsection{Derivation of two-fluid momentum balances and definition of buoyancy closure}
In this subsection, we summarize the results of the mathematically rigorous derivation (i.e., no approximations involved) by \citet{Pahtzetal26} of the macroscopic two-fluid momentum balances from averaging the microscopic equations of motion of the fluid and nearly rigid particles. However, note that their assumption of nearly rigid particles (clarified shortly) is not required for the buoyancy-related analysis we perform afterwards, since considering deformable and/or compressible particles, such as droplets and bubbles, has no effect on the average fluid phase momentum balance and the micromechanical expressions for the quantities therein as long as interfacial mass transfer does not occur \citep{FintziPierson26}.

\citet{Pahtzetal26} defined the microscopic system as follows: The fluid-particle mixture is contained within the domain $\mathbb{V}_\infty=\mathbb{V}_\mathrm{f}[t]\cup\mathbb{V}_\mathrm{s}[t]$ composed of the fluid phase $\mathbb{V}_\mathrm{f}$ and dispersed phase $\mathbb{V}_\mathrm{s}$. The latter is subdivided into individual particles $p$: $\mathbb{V}_\mathrm{s}[t]=\cup_p\mathbb{V}^p[t]$. The corresponding indicator functions $X_\infty[\bm{x}]$, $X_\mathrm{f}[\bm{x},t]$, $X_\mathrm{s}[\bm{x},t]$, and $X^p[\bm{x},t]$, with $\bm{x}\in\mathbb{R}^3$ the spatial coordinate, are defined as being equal to $1$ within the interiors and equal to $0$ outside of $\mathbb{V}_\infty$, $\mathbb{V}_\mathrm{f}$, $\mathbb{V}_\mathrm{s}$, and $\mathbb{V}^p$, respectively. The fluid satisfies general microscopic mass and momentum balance equations:
\begin{align}
 \partial_t\rho+\bm{\nabla}\cdot\rho\bm{u}&=0, \label{MassFluidmic} \\
 \rho\D_t\bm{u}&=\bm{\nabla}\cdot\bm{\sigma}+\rho\bm{g}, \label{MomFluidmic}
\end{align}
where $\D_t\equiv\partial_t+\bm{u}\cdot\bm{\nabla}$. In contrast to before, $\bm{\sigma}$ is a stress tensor associated with an arbitrary fluid flow rheology, and the density $\rho$ is not necessarily constant and therefore distinguished from the previous $\rho_\mathrm{f}$. In particular, the microscopic fields $\rho[\bm{x},t]$ and $\bm{u}[\bm{x},t]$ are not only defined for $\bm{x}\in\mathbb{V}_\mathrm{f}$ but also for $\bm{x}\in\mathbb{V}^p$, where they coincide with a particle $p$'s (not necessarily constant) interior mass density distribution and its velocity $\bm{v}^p_\uparrow+\bm{\omega}^p\times\bm{r}^p$, respectively. Here $\bm{v}^p_\uparrow[t]$ is its centre-of-mass velocity, $\bm{\omega}^p[t]$ its angular velocity, and $\bm{r}^p[\bm{x},t]\equiv\bm{x}-\bm{x}^p[t]$ the coordinate relative to its centre-of-mass location $\bm{x}^p[t]$. A density field $\rho_\ast$ based on the average mass density $M^p/V^p$ of each particle $p$, with $M^p$ its mass and $V^p$ its volume, and a velocity field $\bm{u}_\uparrow$ without particle-rotational contributions are also defined as $\rho_\ast\equiv X_\mathrm{f}\rho+\sum_pX^pM^p/V^p$ and $\bm{u}_\uparrow\equiv X_\mathrm{f}\bm{u}+\sum_pX^p\bm{v}^p_\uparrow$, respectively. The particles are non-Brownian and no interfacial mass transfer is allowed to occur (impermeable boundary conditions), implying that the normal velocity $\bm{u}\cdot\bm{n}^p$ is continuous at the surface $\mathbb{S}^p=\partial\mathbb{V}^p$, where $\bm{n}^p$ is the corresponding unit normal vector pointing outward the particle $p$. They experience the instantaneous, local hydrodynamic force $\bm{F}^p_\mathrm{h}=\int_{\mathbb{S}^p}\bm{n}^p\cdot\bm{\sigma}\d S$, which defines the microscopic hydrodynamic force density field
\begin{equation}
 \bm{f}_\mathrm{h}\equiv\sum\nolimits_p\frac{X^p}{V^p}\bm{F}^p_\mathrm{h}=\sum\nolimits_p\frac{X^p}{V^p}\int_{\mathbb{S}^p}\bm{n}^p\cdot\bm{\sigma}\d S. \label{fh}
\end{equation}
Furthermore, they interact via contacts, where slight transient elastic deformations are permitted (i.e., particles are not fully rigid). However, it is assumed that these do not alter the particles' overall shapes significantly and that the contact area of a contacting particle pair $pq$ can be represented by a single contact point $\bm{x}^{pq}_\mathrm{c}=\bm{x}^{qp}_\mathrm{c}$. The total contact force acting on a particle $p$ is denoted as $\bm{F}^p_\mathrm{c}=\sum_q\bm{F}^{pq}_\mathrm{c}$, resulting from interactions $\bm{F}^{pq}_\mathrm{c}=-\bm{F}^{qp}_\mathrm{c}$ with other particles $q$ (by convention, $\bm{F}^{pp}_\mathrm{c}=0$).

The fluid's and particles' microscopic equations of motion are averaged using an averaging procedure $\langle\cdot\rangle$ that satisfies the following rules \citep{Pahtzetal26}:
\begin{align}
 &\langle cA_1+A_2\rangle=c\langle A_1\rangle+\langle A_2\rangle,&&\langle\bm{\nabla}A\rangle=\bm{\nabla}\langle A\rangle,&&\langle\partial_tA\rangle=\partial_t\langle A\rangle,&&\langle AX_\infty\rangle=\langle A\rangle, \label{AveragingProcedure}
\end{align}
where the $A$-s are arbitrary fields well-defined for $\bm{x}\in\mathbb{V}_\infty$ and $c$ is a constant. In addition, the Reynolds rule,
\begin{equation}
 \langle\langle A_1\rangle A_2\rangle=\langle A_1\rangle\langle A_2\rangle, \label{ReynoldsRule}
\end{equation}
must also be obeyed. It ensures uniqueness of the averaged fields and, when excluding the trivial (and generally unphysical) cases of the identity operator and operators that average over all times and/or space, essentially states that the averaging should include the entire ensemble of microscopic (i.e., fluctuating) states contributing to a macroscopic (i.e., average) state at a given $\bm{x}$ and $t$, while not intermingling different macroscopic states \citep{Pahtzetal26}. At a mathematically exact level, this rule is generally violated for spatio-temporal averaging procedures, since they intermingle generally different macroscopic states from space and time surrounding $\bm{x}$ and $t$. Consider, for example, a vertically-stratified fluid at rest whose properties depend only on the vertical coordinate $z$, where all microscopic states at a given $z$ are identical to each other and the corresponding macroscopic state. Then, spatial averaging that involves the $z$-direction will intermingle different macroscopic states and cause differences between averaged and non-averaged quantities, even though physical intuition dictates that such differences should not occur for a system at rest.

The only spatio-temporal averaging procedures that identically satisfy all the rules in (\ref{AveragingProcedure}) and (\ref{ReynoldsRule}) are the infinite time average in statistically steady systems and infinite averages over spatial coordinates in which the system is homogeneous. If the ergodic hypothesis is true, they are equivalent to ensemble averaging, which also satisfies these rules identically.

Based on $\langle\cdot\rangle$, phase averages and mass-weighted phase averages are also defined \citep{Pahtzetal26}:
\begin{align}
 \langle A\rangle^\mathrm{f}&\equiv\langle X_\mathrm{f}A\rangle/\beta_\mathrm{f},&\langle A\rangle^\mathrm{s}&\equiv\langle X_\mathrm{s}A\rangle/\beta_\mathrm{s},&\langle A\rangle&=\beta_\mathrm{f}\langle A\rangle^\mathrm{f}+\beta_\mathrm{s}\langle A\rangle^\mathrm{s}, \label{PhaseAverages} \\
 \langle A\rangle^\mathrm{f}_{\rho_\ast}&\equiv\langle\rho_\ast A\rangle^\mathrm{f}/\rho_\mathrm{f},&\langle A\rangle^\mathrm{s}_{\rho_\ast}&\equiv\langle\rho_\ast A\rangle^\mathrm{s}/\rho_\mathrm{s},&\langle A\rangle_{\rho_\ast}&\equiv\rho_\mathrm{m}^{-1}\left(\beta_\mathrm{f}\rho_\mathrm{f}\langle A\rangle^\mathrm{f}+\beta_\mathrm{s}\rho_\mathrm{s}\langle A\rangle^\mathrm{s}\right),
\end{align}
where $\beta_\mathrm{f}\equiv\langle X_\mathrm{f}\rangle$ is the fluid phase volume fraction, $\beta_\mathrm{s}\equiv\langle X_\mathrm{s}\rangle$ the dispersed-phase volume fraction, $\rho_\mathrm{f}\equiv\langle\rho_\ast\rangle^\mathrm{f}$ the average fluid density, $\rho_\mathrm{s}\equiv\langle\rho_\ast\rangle^\mathrm{s}$ the average dispersed-phase density, and $\rho_\mathrm{m}\equiv\langle\rho_\ast\rangle=\beta_\mathrm{f}\rho_\mathrm{f}+\beta_\mathrm{s}\rho_\mathrm{s}$ the average mixture density. Note that these definitions are consistent with our prior usage of these quantities. In particular, the non-averaged value of $\rho_\mathrm{f}$ was considered constant and was therefore identical to the averaged value defined here. Furthermore, note that the use of dispersed-phase averages for nearly rigid particles, which are weighted by the indicator function $X^p$ rather than the conventional multivariate delta distribution $\delta[\bm{x}-\bm{x}^p]$, generally lead to particle-rotational terms in translational dispersed-phase balance equations \citep{Pahtzetal26}, see below.

With the above definitions of the microscopic system and averaged quantities, the mathematically exact averaged two-fluid momentum balances can be written as \citep{Pahtzetal26}
\begin{align}
 \beta_\mathrm{f}\rho_\mathrm{f}\D^\mathrm{f}_t\bm{u}_\mathrm{f}&=\bm{\nabla}\cdot\bm{\sigma^\mathrm{f}}+\bm{\nabla}\cdot\bm{\sigma^\mathrm{f}_\mathrm{Re}}+\beta_\mathrm{f}\rho_\mathrm{f}\bm{g}-\beta_\mathrm{s}\langle\bm{f}_\mathrm{h}\rangle^\mathrm{s}, \label{MomFluid} \\
 \beta_\mathrm{s}\rho_\mathrm{s}\D^\mathrm{s}_t\bm{u}_\mathrm{s}&=\bm{\nabla}\cdot\bm{\sigma^\mathrm{s}}+\bm{\nabla}\cdot\bm{\sigma^\mathrm{s}_\mathrm{Re}}+\beta_\mathrm{s}\rho_\mathrm{s}\bm{g}+\beta_\mathrm{s}\langle\bm{f}_\mathrm{h}\rangle^\mathrm{s}, \label{MomSolid}
\end{align}
where $\bm{u}_\mathrm{f}\equiv\langle\bm{u}_\uparrow\rangle_{\rho_\ast}^\mathrm{f}$ is the average fluid phase velocity, $\bm{u}_\mathrm{s}\equiv\langle\bm{u}_\uparrow\rangle_{\rho_\ast}^\mathrm{s}$ the average dispersed-phase velocity without particle-rotational contributions, $\D^\mathrm{f}_t\equiv\partial_t+\bm{u}_\mathrm{f}\cdot\bm{\nabla}$ and $\D^\mathrm{s}_t\equiv\partial_t+\langle\bm{u}\rangle_{\rho_\ast}^\mathrm{s}\cdot\bm{\nabla}$ are the associated material derivatives, and $\bm{\sigma^\mathrm{f}_\mathrm{Re}}\equiv-\beta_\mathrm{f}\rho_\mathrm{f}\left\langle\left(\bm{u}_\uparrow-\bm{u}_\mathrm{f}\right)\left(\bm{u}_\uparrow-\bm{u}_\mathrm{f}\right)\right\rangle_{\rho_\ast}^\mathrm{f}$ and $\bm{\sigma^\mathrm{s}_\mathrm{Re}}\equiv-\beta_\mathrm{s}\rho_\mathrm{s}\left\langle\left(\bm{u}-\langle\bm{u}\rangle_{\rho_\ast}^\mathrm{s}\right)\left(\bm{u}_\uparrow-\bm{u}_\mathrm{s}\right)\right\rangle_{\rho_\ast}^\mathrm{s}$ the associated Reynolds stresses, respectively. Note that, when all particles are sphere-symmetrical, $\langle\bm{u}\rangle_{\rho_\ast}^\mathrm{s}$ can be formally replaced by $\bm{u}_\mathrm{s}$ within the definitions of $\D^\mathrm{s}_t$ and $\bm{\sigma^\mathrm{s}_\mathrm{Re}}$, since their particle-rotational contributions cancel each other out. Furthermore, $\bm{\sigma^\mathrm{f}}\equiv\beta_\mathrm{f}\langle\bm{\sigma}\rangle^\mathrm{f}+\bm{\sigma^\mathrm{f}_\mathrm{s}}$ and $\bm{\sigma^\mathrm{s}}$ are the effective mechanical fluid phase and dispersed-phase stress tensors, respectively. (Note that this notation differs from \citet{Pahtzetal26}, who included the Reynolds stresses $\bm{\sigma^\mathrm{f}_\mathrm{Re}}$ and $\bm{\sigma^\mathrm{s}_\mathrm{Re}}$ within the definitions of $\bm{\sigma^\mathrm{f}}$ and $\bm{\sigma^\mathrm{s}}$, respectively.) The detailed micromechanical expressions for $\bm{\sigma^\mathrm{f}_\mathrm{s}}$ and $\bm{\sigma^\mathrm{s}}$ are provided in Appendix~\ref{StressTensors}, but they are not relevant for the coming analysis. The only relevant aspects are that the pseudo-stress $\bm{\sigma^\mathrm{f}_\mathrm{s}}$ emerges due to the traction by the stress $\bm{\sigma}$ on the particles' surfaces and that it satisfies \citep[Supplementary material of][]{Pahtzetal26}
\begin{equation}
 \bm{\nabla}\cdot\bm{\sigma^\mathrm{f}_\mathrm{s}}=\beta_\mathrm{s}\left\langle\sum\nolimits_p\frac{X^p}{V^p}\int_{\mathbb{S}^p}\bm{n}^p\cdot\langle\bm{\sigma}\rangle^\mathrm{f}\d S\right\rangle^\mathrm{s}+\left\langle\bm{\sigma}^\mathrm{T}\right\rangle^\mathrm{f}\cdot\bm{\nabla}\beta_\mathrm{s}+\bm{\nabla}\cdot\bm{\sigma^\mathrm{f}_\mathrm{s^\prime}}, \label{sigmafs}
\end{equation}
where $\cdot^\mathrm{T}$ denotes the transpose and $\bm{\sigma^\mathrm{f}_\mathrm{s^\prime}}$ is the pseudo-stress that emerges due to the traction by the stress $\bm{\sigma}-\langle\bm{\sigma}\rangle^\mathrm{f}$ on the particles' surfaces (Appendix~\ref{StressTensors}). The latter pseudo-stress therefore vanishes for fluid-particle systems at complete rest due to $\langle\bm{\sigma}\rangle^\mathrm{f}=\bm{\sigma}$.

The averaged two-fluid momentum balances (\ref{MomFluid}) and (\ref{MomSolid}) contain various terms that require closure. However, throughout most of this paper (except in \S\ref{Wellposedness}), we do not employ any closure but the buoyancy closure, defined as the expression for $\beta_\mathrm{s}\langle\bm{f}_\mathrm{B}\rangle^\mathrm{s}$ that separates generalized-drag contributions $\beta_\mathrm{s}\langle\bm{f}_\mathrm{D}\rangle^\mathrm{s}$ from the total interfacial force density $\beta_\mathrm{s}\langle\bm{f}_\mathrm{h}\rangle^\mathrm{s}$:
\begin{equation}
 \beta_\mathrm{s}\langle\bm{f}_\mathrm{h}\rangle^\mathrm{s}=\beta_\mathrm{s}\langle\bm{f}_\mathrm{B}\rangle^\mathrm{s}+\beta_\mathrm{s}\langle\bm{f}_\mathrm{D}\rangle^\mathrm{s}. \label{BuoyancySeparation}
\end{equation}
The goal is to do this in a manner that is mathematically consistent with the exact micromechanical expressions for the other terms in (\ref{MomFluid}) and does not cause disagreements with the particle-resolving simulations and thought experiment presented in \S\ref{PseudoStresses}. In particular, to be consistent with the micromechanical definition of $\bm{f}_\mathrm{h}$, (\ref{fh}), it should be possible to express $\bm{f}_\mathrm{B}$ in an analogous manner as
\begin{equation}
 \bm{f}_\mathrm{B}=\sum\nolimits_p\frac{X^p}{V^p}\left\langle\bm{F}^p_\mathrm{B}\right\rangle=\sum\nolimits_p\frac{X^p}{V^p}\int_{\mathbb{S}^p}\bm{n}^p\cdot\bm{\sigma_\ast}\d S, \label{fB}
\end{equation}
where the average generalized-buoyancy force on a particle $p$, $\langle\bm{F}^p_\mathrm{B}\rangle$, results from the integrated traction by some kind of macroscopic stress $\bm{\sigma_\ast}$ (i.e., a stress that satisfies $\langle\bm{\sigma_\ast}\rangle=\bm{\sigma_\ast}$) over its surface $\mathbb{S}^p$. In two-fluid continuum models, one usually assumes that macroscopic stress divergences such as $\bm{\nabla}\cdot\bm{\sigma_\ast}$ do not vary much within a particle $p$'s volume $\mathbb{V}^p$ \citep{Jackson00}, in which case $\beta_\mathrm{s}\langle\bm{f}_\mathrm{B}\rangle^\mathrm{s}$ approximates as
\begin{equation}
\begin{split}
 \beta_\mathrm{s}\langle\bm{f}_\mathrm{B}\rangle^\mathrm{s}&=\beta_\mathrm{s}\left\langle\sum\nolimits_p\frac{X^p}{V^p}\int_{\mathbb{S}^p}\bm{n}^p\cdot\bm{\sigma_\ast}\d S\right\rangle^\mathrm{s}=\left\langle X_\mathrm{s}\sum\nolimits_p\frac{X^p}{V^p}\int_{\mathbb{V}^p}\bm{\nabla}\cdot\bm{\sigma_\ast}\d V\right\rangle \\
 &=\left\langle\sum\nolimits_p\frac{X^p}{V^p}\int_{\mathbb{V}^p}\bm{\nabla}\cdot\bm{\sigma_\ast}\d V\right\rangle\approx\left\langle\sum\nolimits_pX^p\bm{\nabla}\cdot\bm{\sigma_\ast}\right\rangle=\beta_\mathrm{s}\bm{\nabla}\cdot\bm{\sigma_\ast},
\end{split} \label{fBapprox}
\end{equation}
where we used $X_\mathrm{s}X^p=X^p$ and (\ref{PhaseAverages}). We call this the ``small-particle approximation'' of a buoyancy closure. However, buoyancy closures in the small-particle approximation lead to an inconsistency for fluid-particle systems at complete rest, as shown in the next subsection.

\subsection{Thought experiment: fluid-particle systems at complete rest} \label{SmallParticleApproximation}
For a fluid-particle system at complete rest, $\rho_\mathrm{f}=\rho$ and $\langle\bm{\sigma}\rangle^\mathrm{f}=\bm{\sigma}$ at locations within the fluid phase, $\bm{x}\in V_\mathrm{f}$, and therefore $\bm{\sigma^\mathrm{f}_\mathrm{s^\prime}}=0$. Likewise, velocities and generalized-drag forces vanish due to the absence of motion. Then, the generalized-buoyancy force density $\beta_\mathrm{s}\langle\bm{f}_\mathrm{B}\rangle^\mathrm{s}$ must be equal to the hydrodynamic force density $\beta_\mathrm{s}\langle\bm{f}_\mathrm{h}\rangle^\mathrm{s}$, which implies $\bm{\sigma_\ast}=\bm{\sigma}$ for $\bm{x}\in V_\mathrm{f}$ from comparing (\ref{fB}) with (\ref{fh}). Using $\bm{\sigma^\mathrm{f}}\equiv\beta_\mathrm{f}\langle\bm{\sigma}\rangle^\mathrm{f}+\bm{\sigma^\mathrm{f}_\mathrm{s}}$ and (\ref{sigmafs}), it follows that the macroscopic fluid phase momentum balance (\ref{MomFluid}) reduces to the stationary microscopic fluid phase momentum balance,
\begin{equation}
 X_\mathrm{f}(\bm{\nabla}\cdot\bm{\sigma}+\rho\bm{g})=0, \label{MomFluidmicstat}
\end{equation}
as required. However, this consistency property is lost when using a buoyancy closure in the small-particle approximation. Then, (\ref{MomFluidmicstat}) is recovered generally only for fluids where $\rho$ does not vary with $\bm{x}\in V_\mathrm{f}$. In particular, a particle bed immersed in a density-stratified fluid at rest would violate this property. This is not just a minor nuisance but is indirect evidence of a more fundamental issue. As we show in \S\ref{Wellposedness}, it is precisely the small-particle approximation that makes buoyancy cause ill-posedness of some two-fluid continuum models.

\subsection{Previously proposed buoyancy closures}
When substituting (\ref{sigmafs}) via $\bm{\sigma^\mathrm{f}}\equiv\beta_\mathrm{f}\langle\bm{\sigma}\rangle^\mathrm{f}+\bm{\sigma^\mathrm{f}_\mathrm{s}}$ in the fluid phase momentum balance (\ref{MomFluid}) and using the definition of the hydrodynamic microscopic force density $\bm{f}_\mathrm{h}$, (\ref{fh}), one obtains
\begin{equation}
 \beta_\mathrm{f}\rho_\mathrm{f}\D^\mathrm{f}_t\bm{u}_\mathrm{f}=\beta_\mathrm{f}\bm{\nabla}\cdot\langle\bm{\sigma}\rangle^\mathrm{f}+\bm{\nabla}\cdot\left(\bm{\sigma^\mathrm{f}_\mathrm{s^\prime}}+\bm{\sigma^\mathrm{f}_\mathrm{Re}}\right)+\beta_\mathrm{f}\rho_\mathrm{f}\bm{g}-\beta_\mathrm{s}\left\langle\sum\nolimits_p\frac{X^p}{V^p}\int_{\mathbb{S}^p}\bm{n}^p\cdot\left(\bm{\sigma}-\langle\bm{\sigma}\rangle^\mathrm{f}\right)\d S\right\rangle^\mathrm{s}. \label{MomFluidZhang}
\end{equation}
This mathematical structure led \citet{Zhangetal07b} to indirectly, and later others \citep[e.g.,][]{Wangetal24} to directly, suggest that the last term on the right-hand side of (\ref{MomFluidZhang}) is the negative generalized-drag force density, $-\beta_\mathrm{s}\langle\bm{f}_\mathrm{D}\rangle^\mathrm{s}$, and that
\begin{equation}
 \text{Zhang et al. (2007):}\quad\beta_\mathrm{s}\langle\bm{f}_\mathrm{B}\rangle^\mathrm{s}=\beta_\mathrm{s}\left\langle\sum\nolimits_p\frac{X^p}{V^p}\int_{\mathbb{S}^p}\bm{n}^p\cdot\langle\bm{\sigma}\rangle^\mathrm{f}\d S\right\rangle^\mathrm{s} \label{BuoyancyZhang}
\end{equation}
is therefore the generalized-buoyancy force density. This closure expression implies that $\langle\bm{\sigma}\rangle^\mathrm{f}$ is the background flow stress responsible for buoyancy, that is, $\bm{\sigma_\ast}=\langle\bm{\sigma}\rangle^\mathrm{f}$ in (\ref{fB}).

\citet{Ishii75} proposed an alternative closure. They started with a different form of the fluid phase momentum balance,
\begin{equation}
 \beta_\mathrm{f}\rho_\mathrm{f}\D^\mathrm{f}_t\bm{u}_\mathrm{f}=\beta_\mathrm{f}\bm{\nabla}\cdot\langle\bm{\sigma}\rangle^\mathrm{f}+\bm{\nabla}\cdot\bm{\sigma^\mathrm{f}_\mathrm{Re}}+\beta_\mathrm{f}\rho_\mathrm{f}\bm{g}+\left\langle\left(\bm{\sigma}^\mathrm{T}-\left\langle\bm{\sigma}^\mathrm{T}\right\rangle^\mathrm{f}\right)\cdot\bm{\nabla}X_\mathrm{s}\right\rangle, \label{MomFluidDrew}
\end{equation}
which follows from (\ref{MomFluid}) and the identity $\beta_\mathrm{s}\langle\bm{f}_\mathrm{h}\rangle^\mathrm{s}=\bm{\nabla}\cdot\bm{\sigma}^\mathrm{f}_\mathrm{s}-\left\langle\bm{\sigma}^\mathrm{T}\cdot\bm{\nabla}X_\mathrm{s}\right\rangle$ \citep{Pahtzetal26}. They then interpreted the last term on the right-hand side of (\ref{MomFluidDrew}) as the negative generalized-drag force density, $-\beta_\mathrm{s}\langle\bm{f}_\mathrm{D}\rangle^\mathrm{s}=\langle(\bm{\sigma}^\mathrm{T}-\langle\bm{\sigma}^\mathrm{T}\rangle^\mathrm{f})\cdot\bm{\nabla}X_\mathrm{s}\rangle$. From comparison with (\ref{MomFluidZhang}), one can see that this interpretation is equivalent to assuming
\begin{equation}
 \text{Ishii (1975):}\quad\beta_\mathrm{s}\langle\bm{f}_\mathrm{B}\rangle^\mathrm{s}=\beta_\mathrm{s}\left\langle\sum\nolimits_p\frac{X^p}{V^p}\int_{\mathbb{S}^p}\bm{n}^p\cdot\langle\bm{\sigma}\rangle^\mathrm{f}\d S\right\rangle^\mathrm{s}+\bm{\nabla}\cdot\bm{\sigma^\mathrm{f}_\mathrm{s^\prime}}. \label{BuoyancyDrew}
\end{equation}
However, this closure does not allow the identification of a background flow stress $\bm{\sigma_\ast}$ that is responsible for buoyancy. This is not surprising, since \citet{Ishii75}, as it was the fashion at the time \citep[e.g.,][]{Drew83}, falsely treated the terms $\langle\bm{\sigma}^\mathrm{T}\cdot\bm{\nabla}X_\mathrm{s}\rangle$, $\langle\langle\bm{\sigma}^\mathrm{T}\rangle^\mathrm{f}\cdot\bm{\nabla}X_\mathrm{s}\rangle$, and $\langle(\bm{\sigma}^\mathrm{T}-\langle\bm{\sigma}^\mathrm{T}\rangle^\mathrm{f})\cdot\bm{\nabla}X_\mathrm{s}\rangle$ as actual fluid-particle interaction force densities, even though their micromechanical definitions do not take the particles' surface geometries into account. In addition, even in the small-particle approximation, in which case (\ref{BuoyancyDrew}) becomes $\beta_\mathrm{s}\langle\bm{f}_\mathrm{B}\rangle^\mathrm{s}\approx\beta_\mathrm{s}\bm{\nabla}\cdot\langle\bm{\sigma}\rangle^\mathrm{f}+\bm{\nabla}\cdot\bm{\sigma^\mathrm{f}_\mathrm{s^\prime}}$, it is still not possible to identify a background flow stress $\bm{\sigma_\ast}$, since this would require an expression of the form $\beta_\mathrm{s}\langle\bm{f}_\mathrm{B}\rangle^\mathrm{s}\approx\beta_\mathrm{s}\bm{\nabla}\cdot\bm{\sigma_\ast}$ as shown in (\ref{fBapprox}). In other words, the closure (\ref{BuoyancyDrew}) fails at a conceptual level.

Two further buoyancy closures based on the small-particle approximation have been put forward. \citet{RevilBaudardChauchat13} essentially proposed 
\begin{equation}
 \text{Revil-Baudard \& Chauchat (2013):}\quad\beta_\mathrm{s}\langle\bm{f}_\mathrm{B}\rangle^\mathrm{s}=\beta_\mathrm{s}\bm{\nabla}\cdot\bm{\sigma^\mathrm{f}}, \label{BuoyancyChauchat}
\end{equation}
since they assumed that a mixture-viscosity-based shear stress, which micromechanically originates from the stress $\bm{\sigma^\mathrm{f}_\mathrm{s^\prime}}$ \citep{Jackson97}, contributes to the background flow stress responsible for buoyancy. Furthermore, \citet{AndersonJackson67} and \citet{Jackson00}, though their micromechanical expressions for the stress tensor $\bm{\sigma^\mathrm{f}}$ were based on different levels of approximations, proposed the same buoyancy closure, henceforth called the ``Jackson closure'':
\begin{equation}
 \text{Jackson closure:}\quad\beta_\mathrm{s}\langle\bm{f}_\mathrm{B}\rangle^\mathrm{s}=\beta_\mathrm{s}\bm{\nabla}\cdot\left(\bm{\sigma^\mathrm{f}}+\bm{\sigma^\mathrm{f}_\mathrm{Re}}\right). \label{BuoyancyJackson}
\end{equation}
Noting that $\bm{\sigma^\mathrm{f}}\equiv\beta_\mathrm{f}\langle\bm{\sigma}\rangle^\mathrm{f}+\bm{\sigma^\mathrm{f}_\mathrm{s}}\approx\langle\bm{\sigma}\rangle^\mathrm{f}+\bm{\sigma^\mathrm{f}_\mathrm{s^\prime}}$ in the small-particle approximation, (\ref{BuoyancyChauchat}) and (\ref{BuoyancyJackson}) differ from $\beta_\mathrm{s}\langle\bm{f}_\mathrm{B}\rangle^\mathrm{s}\approx\beta_\mathrm{s}\bm{\nabla}\cdot\langle\bm{\sigma}\rangle^\mathrm{f}$, the small-particle approximation of (\ref{BuoyancyZhang}), in the additional inclusion of $\bm{\sigma^\mathrm{f}_\mathrm{s^\prime}}$ and $\bm{\sigma^\mathrm{f}_\mathrm{s^\prime}}+\bm{\sigma^\mathrm{f}_\mathrm{Re}}$, respectively, within the divergence operator. We comment on these closures in the next subsection.

\subsection{Physically inspired buoyancy criterion} \label{BuoyancyCriterion}
We propose that a physically meaningful buoyancy closure should satisfy the criterion 
\begin{equation}
 \bm{\nabla}\cdot\bm{\sigma^\mathrm{f}}=\beta_\mathrm{f}\bm{\nabla}\cdot\bm{\hat\sigma^\mathrm{f}}+\beta_\mathrm{s}\langle\bm{f}_\mathrm{B}\rangle^\mathrm{s}, \label{Criterion}
\end{equation}
with a generalized-buoyancy force density that respects the geometry of the particles and therefore the mathematical structure (\ref{fB}):
\begin{equation}
 \beta_\mathrm{s}\langle\bm{f}_\mathrm{B}\rangle^\mathrm{s}=\beta_\mathrm{s}\left\langle\sum\nolimits_p\frac{X^p}{V^p}\int_{\mathbb{S}^p}\bm{n}^p\cdot\bm{\hat\sigma^\mathrm{f}}\d S\right\rangle^\mathrm{s}. \label{BuoyancyHypothesis}
\end{equation}
When integrating (\ref{Criterion}) over an infinitesimally small mixture volume $\d V$, it states that the net flux $\d V\bm{\nabla}\cdot\bm{\sigma^\mathrm{f}}$ of average fluid momentum into $\d V$ decomposes into the net flux $\d V_\mathrm{f}\bm{\nabla}\cdot\bm{\hat\sigma^\mathrm{f}}$ of average fluid momentum into $\d V_\mathrm{f}\equiv\beta_\mathrm{f}\d V$, the average fluid phase portion of $\d V$, and the net flux $\d V_\mathrm{s}\langle\bm{f}_\mathrm{B}\rangle^\mathrm{s}$ of average fluid momentum into $\d V_\mathrm{s}\equiv\beta_\mathrm{s}\d V$, the average dispersed-phase portion of $\d V$. Importantly, both momentum flux contributions must be caused by the same stress $\bm{\hat\sigma^\mathrm{f}}$, which can then be interpreted as the average background flow stress.

Using $\beta_\mathrm{f}\langle\rho\D_t\bm{u}\rangle^\mathrm{f}=\beta_\mathrm{f}\rho_\mathrm{f}\D^\mathrm{f}_t\bm{u}_\mathrm{f}-\bm{\nabla}\cdot\bm{\sigma^\mathrm{f}_\mathrm{Re}}$ \citep{Pahtzetal26} and (\ref{BuoyancySeparation}), the criterion (\ref{Criterion}) allows transforming the fluid phase momentum balance (\ref{MomFluid}) into the form
\begin{equation}
 \beta_\mathrm{f}\langle\rho\D_t\bm{u}\rangle^\mathrm{f}=\beta_\mathrm{f}\bm{\nabla}\cdot\bm{\hat\sigma^\mathrm{f}}+\beta_\mathrm{f}\rho_\mathrm{f}\bm{g}-\beta_\mathrm{s}\langle\bm{f}_\mathrm{D}\rangle^\mathrm{s}. \label{MomFluidHypothesis}
\end{equation}
Its mathematical structure differs from the fluid phase momentum balance resulting from the closure by \citet{Zhangetal07b}, (\ref{MomFluidZhang}), in a single aspect: (\ref{MomFluidHypothesis}) does not contain a structural analogue to the term $\bm{\nabla}\cdot\bm{\sigma^\mathrm{f}_\mathrm{s^\prime}}$ in (\ref{MomFluidZhang}) on its right-hand side. When integrated over an infinitesimally small mixture volume $\d V$, (\ref{MomFluidHypothesis}) states that the average temporal change $\d V_\mathrm{f}\langle\rho\D_t\bm{u}\rangle^\mathrm{f}$ of fluid momentum within the pure-fluid volume $\d V_\mathrm{f}$ is due to the net flux $\d V_\mathrm{f}\bm{\nabla}\cdot\bm{\hat\sigma^\mathrm{f}}$ of average fluid momentum into $\d V_\mathrm{f}$ and the body force, consisting of the gravitational and generalized-drag force terms. This is completely analogous to the microscopic fluid phase momentum balance (\ref{MomFluidmic}) integrated over an infinitesimally small pure-fluid volume $\d V_\mathrm{f}$, where the temporal change $\d V_\mathrm{f}\rho\D_t\bm{u}$ of fluid momentum within $\d V_\mathrm{f}$ is due to the net flux $\d V_\mathrm{f}\bm{\nabla}\cdot\bm{\sigma}$ of fluid momentum into $\d V_\mathrm{f}$ and the body force, consisting of only the gravitational term in this case. By contrast, such an analogy would not exist if there were mechanical stress divergence terms in the fluid phase momentum balance that were not being multiplied by $\beta_\mathrm{f}$, such as the term $\bm{\nabla}\cdot\bm{\sigma^\mathrm{f}_\mathrm{s^\prime}}$ in (\ref{MomFluidZhang}). Such terms would then still encode net fluxes of average fluid momentum into the mixture, as opposed to only the average fluid phase portion of the mixture, and therefore into the particles. It would mean that the net momentum transfer from the fluid to the particles would not be completely ``outsourced'' to the interfacial force density terms, contrary to what we propose a buoyancy closure should actually achieve. In fact, in (\ref{MomFluidHypothesis}), fluid-to-particle momentum transfer is entirely encoded in the generalized-drag force density $\beta_\mathrm{s}\langle\bm{f}_\mathrm{D}\rangle^\mathrm{s}$.

The key questions now are: Does a stress $\bm{\hat\sigma^\mathrm{f}}$ that satisfies the criterion (\ref{Criterion}) exist? And, if it exists, what is its micromechanical expression? These questions will be positively answered in \S\ref{NewClosure}. At the present stage, one can already infer from comparing (\ref{MomFluidHypothesis}) with (\ref{MomFluidZhang}) that $\bm{\hat\sigma^\mathrm{f}}=\langle\bm{\sigma}\rangle^\mathrm{f}$ for fluid-particle systems at complete rest due to $\bm{\sigma^\mathrm{f}_\mathrm{s^\prime}}=0$. By contrast, generally $\bm{\sigma^\mathrm{f}}\ne\langle\bm{\sigma}\rangle^\mathrm{f}$ for such systems due to (\ref{sigmafs}). Furthermore, from (\ref{Criterion}), it can already be seen that $\bm{\hat\sigma^\mathrm{f}}\approx\bm{\sigma^\mathrm{f}}$ when applying the small-particle approximation $\beta_\mathrm{s}\langle\bm{f}_\mathrm{B}\rangle^\mathrm{s}\approx\beta_\mathrm{s}\bm{\nabla}\cdot\bm{\hat\sigma^\mathrm{f}}$. As a consequence, (\ref{BuoyancyHypothesis}) reduces to the closure by \citet{RevilBaudardChauchat13}, (\ref{BuoyancyChauchat}), in this approximation.

Finally, it is worth noting that, when altering the criterion (\ref{Criterion}) in a manner that leads to an alternative version of (\ref{MomFluidHypothesis}) in which the average temporal change of fluid momentum per unit volume, $\beta_\mathrm{f}\langle\rho\D_t\bm{u}\rangle^\mathrm{f}$, on its left-hand side is replaced by the temporal change of the average fluid momentum per unit volume, $\beta_\mathrm{f}\rho_\mathrm{f}\D^\mathrm{f}_t\bm{u}_\mathrm{f}$, one would obtain the Jackson closure, (\ref{BuoyancyJackson}), in the small-particle approximation, which additionally includes the Reynolds stress $\bm{\sigma^\mathrm{f}_\mathrm{Re}}$ within the divergence operator. However, this appears inconsistent with the micromechanical expressions for $\beta_\mathrm{s}\langle\bm{f}_\mathrm{h}\rangle^\mathrm{s}$ and $\beta_\mathrm{s}\langle\bm{f}_\mathrm{B}\rangle^\mathrm{s}$, based on (\ref{fh}) and (\ref{fB}), respectively, since they refer to the average force per unit volume acting on actual particles rather than the force per unit volume acting on average particles. This statement and the physical rationale put forward in this subsection are strongly supported by the particle-resolving numerical simulations and thought experiment presented in the next section.

\section{Contributions of pseudo-stresses to buoyancy} \label{PseudoStresses}
In this section, we use particle-resolving simulations (\S\ref{Simulations}) and a thought experiment (\S\ref{StressFluidParticleInteractions}) to test existing buoyancy closures. The results of this section are summarized in \S\ref{ConclusionPseudoStresses}.
\subsection{Particle-resolving simulations} \label{Simulations}
We use the same numerical model and code as \citet{Biegertetal17}, which have undergone thorough testing and calibration in previous studies, including grid convergence tests and others \citep{Biegert18,Yaoetal22,Zhuetal22}. The model is based on the immersed boundary method after \citet{Uhlmann05}, where each particle $p$'s interior is replaced by pseudo-fluid, while a Lagrangian surface force density $\bm{\tilde f}^p_\mathrm{IBM}$ enforces no-slip boundary conditions at its surface $\mathbb{S}^p$. This surface is modelled as a very thin volume layer, the so-called Lagrangian layer \citep{Tschisgaleetal17}, extending minimally both into the interior and exterior of the particle $p$. A smooth volumetric force density derived from $\bm{\tilde f}^p_\mathrm{IBM}$ also enters the Navier-Stokes equations as an additional body force term. These governing equations are then numerically integrated, on a uniform rectangular grid of length $\Delta=0.1R$, using a third-order low-storage Runge-Kutta scheme for time advancement and a finite-difference approach for spatial discretization. The pressure field is solved using a direct solver based on the fast Fourier transform. To integrate the equations governing the particles' motion, the Runge-Kutta scheme subdivides the three-step fluid integration procedure into a total of $15$ substeps per fluid time step. The hydrodynamic force acting on a particle $p$ is calculated as
\begin{equation}
 \bm{F}^p_\mathrm{h}=\d_t\int_{\mathbb{V}^p}\rho_\mathrm{f}\bm{u}\d V-\int_{\mathbb{S}^p}\bm{\tilde f}^p_\mathrm{IBM}\d S-\rho_\mathrm{f}\bm{g}V^p, \label{FhIBM}
\end{equation}
which exploits that the instantaneous, local fluid velocity $\bm{u}$ is solved in both the exterior and interior of the particles. Furthermore, contact forces are calculated using the soft-sphere model proposed by \citet{Biegertetal17}. When the distance between the surfaces of two approaching particles becomes sufficiently small, the intervening fluid is squeezed out of the gap. Once the gap width falls below approximately $2\Delta$, this process can no longer be resolved by the fluid grid. Therefore, a lubrication model, equation~(9) of \citet{Biegertetal17}, is employed to account for the unresolved hydrodynamic interactions. This model is also applied to particles rebounding after collision, when fluid is drawn back into the gap. The lubrication force is dissipative, since it always acts in the direction opposite to the relative velocity. Besides, the adaptive collision time Model proposed by \citet{KempeFrohlich12a} is employed to account for normal contact interactions. It stretches the collision process in time to match the time step of the fluid solver.

\subsubsection{Sediment transport driven by viscous flow} \label{SedimentTransport}
\begin{figure}
\centering
 \includegraphics{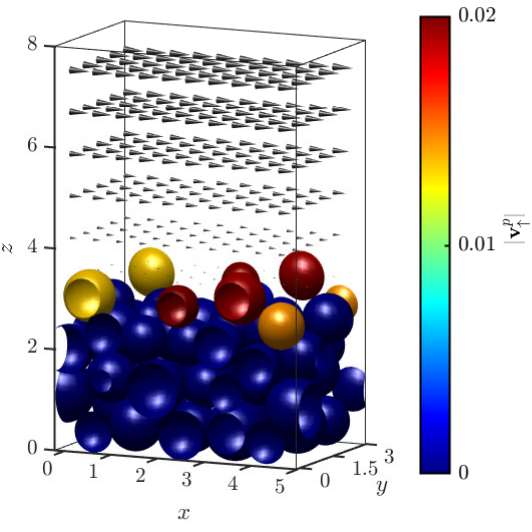}
\caption{Snapshot of a particle-resolving simulation of statistically steady, uniform sediment transport driven by viscous flow.}
\label{Sketch}
\end{figure}

We use our previously reported simulation \citep{Pahtzetal26} of statistically steady, uniform sediment transport driven by viscous flow (figure~\ref{Sketch}).
This system is characterized by a negligible generalized-drag force density $\beta_\mathrm{s}\langle\bm{f}_\mathrm{D}\rangle^\mathrm{s}$ in the bed-normal direction (explained shortly), a property that allows direct access to the generalized-buoyancy force density $\beta_\mathrm{s}\langle\bm{f}_\mathrm{B}\rangle^\mathrm{s}$, which is otherwise difficult to isolate.

The gravitational acceleration $\bm{g}$ is driving both the liquid flow and granular particles down a small slope of $3^\circ$. Physical quantities are measured in natural units of $\rho_\mathrm{f}$, $(\rho_\mathrm{s}/\rho_\mathrm{f}-1)|\bm{g}|$, and $2R$. The simulation was conducted for $\rho_\mathrm{s}=2.65$ and a high viscosity $\eta_\mathrm{f}=0.36$, yielding a weakly inertial flow regime with a bulk flow Reynolds number $\Re\equiv\rho_\mathrm{f}\max(u_{\mathrm{f}x})H/\eta_\mathrm{f}\approx22.4$, where $H$ is the extension of the vertical domain. The flow direction is $x$, the direction normal to the bed is $z$ (henceforth called the ``vertical'' direction for simplicity), and the lateral direction is $y$. The system dimensions are $(L_x,L_y,H)=(5,3,8)$, and $50$ identical, spherical particles are considered, forming a bed of approximately four particle layers, with the top layer being the most mobile (figure~\ref{Sketch}). Here $L_x$ and $L_y$ denote the length and width, respectively, of the $(x,y)$-domain periodic in both $x$ and $y$, while no-slip and free-slip boundary conditions are imposed at the bottom ($z=0$) and top ($z=H$) walls, respectively, of the domain to mimic a rigid and free surface.

The fluid's and particles' equation of motion are averaged using the following averaging procedure:
\begin{equation}
 \langle A\rangle[z_n]\equiv\frac{1}{N_TL_xL_y\Delta z}\sum_{k=1}^{N_T}\int_{z_n-\Delta z/2}^{z_n+\Delta z/2}\int_0^{L_y}\int_0^{L_x}A[\bm{x},t_k]\d x\d y\d z, \label{AveragingDNSDEM}
\end{equation}
which satisfies the rules in (\ref{AveragingProcedure}). Due to statistical bed-tangential homogeneity ($\partial_x=\partial_y=0$), the averaging takes place over the entire periodic bed-tangential domain, $x\in[0,L_x)$ and $y\in[0,L_y)$, and, due to statistical steadiness, over a sufficient number $N_T=268$ of timewise \textit{well-separated} instants $t_k$ to ensure $\partial_t\approx0$. Furthermore, $\Delta z=0.1$ is the constant extent of the intervals with centers $z_n$ that discretize the bed-normal domain $z\in(0,H)$. It is twice the simulation's grid length $\Delta=0.05$ and sufficiently small to ensure that (\ref{AveragingDNSDEM}) approximately satisfies the Reynolds rule (\ref{ReynoldsRule}). By contrast, since the macroscopic flow can change very rapidly over the particle diameter $2R$ in the $z$-direction, a value of $\Delta z$ or the coarse-graining width of the order of $2R$, as used in some previous studies \citep[e.g.,][]{Fryetal24}, would be too large. For reasons described by \citet{Pahtzetal26}, it is also undesirable to use coarse-graining kernels without a compact support, such as the often used Gaussian kernels \citep{Goldhirsch10}.

Due to $\partial_x=\partial_y=\partial_t=0$, the average fluid phase momentum balance (\ref{MomFluid}) in the $x$- and $z$-directions, respectively, becomes:
\begin{align}
 \frac{\d}{\d z}\left(\sigma^\mathrm{f}_{zx}+\sigma^\mathrm{f}_{\mathrm{Re}zx}\right)&=-\beta_\mathrm{f}\rho_\mathrm{f}g_x+\beta_\mathrm{s}\langle f_{\mathrm{h}x}\rangle^\mathrm{s}, \label{SigmaFluidSteadyzx} \\
 \frac{\d}{\d z}\left(\sigma^\mathrm{f}_{zz}+\sigma^\mathrm{f}_{\mathrm{Re}zz}\right)&=-\beta_\mathrm{f}\rho_\mathrm{f}g_z+\beta_\mathrm{s}\langle f_{\mathrm{h}z}\rangle^\mathrm{s}. \label{SigmaFluidSteadyzz}
\end{align}
Almost all coarse-grained quantities therein are computed directly as described by \citet{Pahtzetal26}. The only exception is $\bm{\sigma^\mathrm{f}_\mathrm{s}}$ within $\bm{\sigma^\mathrm{f}}\equiv\beta_\mathrm{f}\langle\bm{\sigma}\rangle^\mathrm{f}+\bm{\sigma^\mathrm{f}_\mathrm{s}}$, which is calculated indirectly from the other quantities via integrating the right-hand sides of (\ref{SigmaFluidSteadyzx}) and (\ref{SigmaFluidSteadyzz}) to minimize potential numerical inaccuracies, since its direct calculation is the least accurately determinable term for the immersed boundary method \citep{Pahtzetal26}.

\begin{figure}
\centering
 \includegraphics[width=\textwidth]{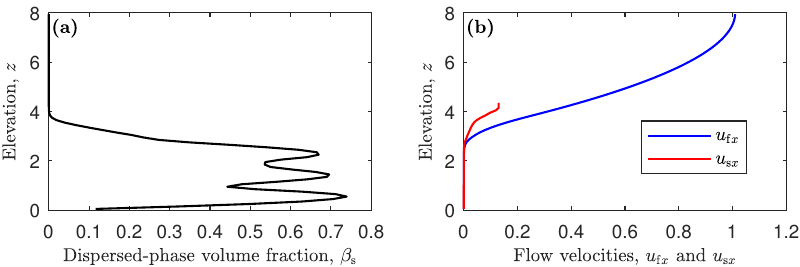}
\caption{Vertical profiles of (a) the dispersed-phase volume fraction $\beta_\mathrm{s}$ and (b) the fluid-phase-averaged streamwise flow velocity $u_{\mathrm{f}x}$ and dispersed-phase-averaged streamwise flow velocity $u_{\mathrm{s}x}$, obtained from a particle-resolving simulation of statistically steady, uniform viscous sediment transport.}
\label{ConcentrationVelocityProfiles}
\end{figure}

Figure~\ref{ConcentrationVelocityProfiles} shows the vertical profiles of the dispersed-phase volume fraction $\beta_\mathrm{s}$, fluid-phase-averaged streamwise flow velocity $u_{\mathrm{f}x}$, and dispersed-phase-averaged streamwise flow velocity $u_{\mathrm{s}x}$.
All three kinds of profiles are qualitatively consistent with expectations from existing sediment transport simulations \citep[e.g.,][]{Maurinetal15,PahtzDuran18a} and highly-resolved experiments \citep{NiCapart18}. In particular, the profiles of $\beta_\mathrm{s}$ indicate layering within the sediment bed and a bed-transport-layer interface at $z\approx3$ (figure~\ref{ConcentrationVelocityProfiles}(a)).

\begin{figure}
\centering
 \includegraphics[width=\textwidth]{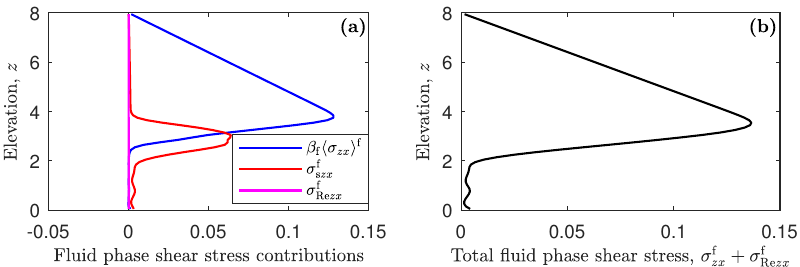}
\caption{(a) Fluid phase shear stress contributions obtained from a particle-resolving simulation of statistically steady, uniform viscous sediment transport. (b) The sum of all contributions results in the total fluid phase shear stress $\sigma^\mathrm{f}_{zx}+\sigma_{\mathrm{Re}zx}^\mathrm{f}=\beta_\mathrm{f}\langle\sigma_{zx}\rangle^\mathrm{f}+\sigma_{\mathrm{s}zx}^\mathrm{f}+\sigma_{\mathrm{Re}zx}^\mathrm{f}$.}
\label{Fig_SigmaProfileszx}
\end{figure}

\begin{figure}
\centering
 \includegraphics[width=\textwidth]{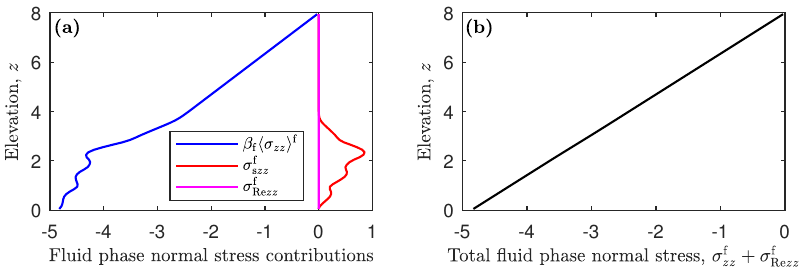}
\caption{(a) Fluid phase normal stress contributions obtained from a particle-resolving simulation of statistically steady, uniform viscous sediment transport. (b) The sum of all contributions results in the total fluid phase normal stress $\sigma^\mathrm{f}_{zz}+\sigma_{\mathrm{Re}zz}^\mathrm{f}=\beta_\mathrm{f}\langle\sigma_{zz}\rangle^\mathrm{f}+\sigma_{\mathrm{s}zz}^\mathrm{f}+\sigma_{\mathrm{Re}zz}^\mathrm{f}$, which exhibits a nearly hydrostatic profile, $\sigma^\mathrm{f}_{zz}+\sigma_{\mathrm{Re}zz}^\mathrm{f}\approx\rho_\mathrm{f}(H-z)g_z$.}
\label{Fig_SigmaProfileszz}
\end{figure}

Figures~\ref{Fig_SigmaProfileszx} and \ref{Fig_SigmaProfileszz} show the vertical profiles of the total fluid phase shear ($\sigma^\mathrm{f}_{zx}+\sigma^\mathrm{f}_{\mathrm{Re}zx}$) and normal ($\sigma^\mathrm{f}_{zz}+\sigma^\mathrm{f}_{\mathrm{Re}zz}$) stresses, respectively, and their constituting contributions based on $\bm{\sigma^\mathrm{f}}\equiv\beta_\mathrm{f}\langle\bm{\sigma}\rangle^\mathrm{f}+\bm{\sigma^\mathrm{f}_\mathrm{s}}$.
In particular, it can be seen that the Reynolds stress $\bm{\sigma^\mathrm{f}_\mathrm{Re}}$ is negligible (figures~\ref{Fig_SigmaProfileszx}(a) and \ref{Fig_SigmaProfileszz}(a)), as expected, and that the vertical profiles of $\sigma^\mathrm{f}_{zz}+\sigma^\mathrm{f}_{\mathrm{Re}zz}$ are approximately hydrostatic (figure~\ref{Fig_SigmaProfileszz}(b)), $\sigma^\mathrm{f}_{zz}+\sigma^\mathrm{f}_{\mathrm{Re}zz}\approx\rho_\mathrm{f}(H-z)g_z$, as was also reported by \citet{Fryetal24} for their simulations of viscous sediment transport.

\begin{figure}
\centering
 \includegraphics[width=\textwidth]{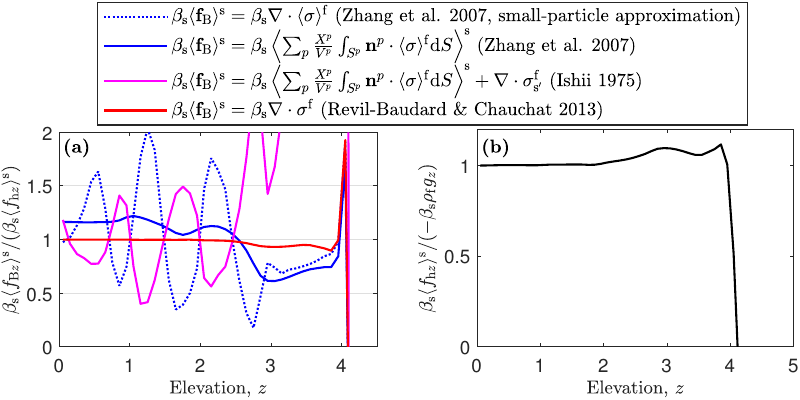}
\caption{(a) Test of the buoyancy closures by \citet{Zhangetal07b}, (\ref{BuoyancyZhang}), \citet{Ishii75}, (\ref{BuoyancyDrew}), and \citet{RevilBaudardChauchat13}, (\ref{BuoyancyChauchat}), against data from a particle-resolving simulation of statistically steady, uniform viscous sediment transport. If lift forces are not significant, as indicated by $\beta_\mathrm{s}\langle f_{\mathrm{h}z}\rangle^\mathrm{s}/(-\beta_\mathrm{s}\rho_\mathrm{f}g_z)\approx1$ in (b), one expects $\beta_\mathrm{s}\langle f_{\mathrm{B}z}\rangle^\mathrm{s}/(\beta_\mathrm{s}\langle f_{\mathrm{h}z}\rangle^\mathrm{s})\approx1$ in (a). This is, indeed, the case for the closure by \citet{RevilBaudardChauchat13}, but not for the other closures. The peaks around $z=4$ likely have a numerical origin due to insufficient data for statistical averaging when $\beta_\mathrm{s}$ is very small (figure~\ref{ConcentrationVelocityProfiles}(a)).}
\label{Fig_BuoyancyClosures}
\end{figure}

Due to the weakly-inertial flow conditions (low Reynolds and Stokes numbers), the instantaneous hydrodynamic forces acting on the transported particles are approximately given by the background-flow (buoyancy) and drag forces in the MRG equation (\ref{MRG}), and lubrication forces due to fluid-mediated particle interactions. However, as already mentioned, short-range lubrication forces are not resolved by the numerical model, but explicitly accounted for via contact force corrections, i.e., they contribute to the contact stress $\bm{\sigma^\mathrm{s}}$ rather than the hydrodynamic force density $\beta_\mathrm{s}\langle\bm{f}_\mathrm{h}\rangle^\mathrm{s}$ in (\ref{MomFluid}) and (\ref{MomSolid}). Therefore, in the vertical direction, where $u_{\mathrm{f}z}=u_{\mathrm{s}z}=0$ due to the steady state condition, drag forces on transported particles cancel out on average and only the buoyancy force $F_{\mathrm{B}z}$ remains, which is approximately equal to $-(4/3)\pi R^3\rho_\mathrm{f}g_z$ \citep{MaxeyRiley83}. Hence, we expect
\begin{equation}
 \beta_\mathrm{s}\langle f_{\mathrm{h}z}\rangle^\mathrm{s}\approx\beta_\mathrm{s}\langle f_{\mathrm{B}z}\rangle^\mathrm{s}\approx-\beta_\mathrm{s}\rho_\mathrm{f}g_z. \label{FBz}
\end{equation}
In figure~\ref{Fig_BuoyancyClosures}, we use this expectation to test the buoyancy closures by \citet{Zhangetal07b}, (\ref{BuoyancyZhang}), \citet{Ishii75}, (\ref{BuoyancyDrew}), and \citet{RevilBaudardChauchat13}, (\ref{BuoyancyChauchat}), against the simulation data.
It can be seen that these data clearly support the latter closure. It means that the full effective mechanical fluid phase stress $\bm{\sigma^\mathrm{f}}$, rather than only the fluid-phase-averaged stress $\langle\bm{\sigma}\rangle^\mathrm{f}$, is the background flow stress responsible for buoyancy. This result is further confirmed by the thought experiment in \S\ref{StressFluidParticleInteractions}. Note that these simulation data are unable to distinguish between the closure by \citet{RevilBaudardChauchat13}, (\ref{BuoyancyChauchat}), and the Jackson closure, (\ref{BuoyancyJackson}), due to $\sigma^\mathrm{f}_{\mathrm{Re}zz}\approx0$.

\subsubsection{``Horizontal settling''} \label{HorizontalSettling}
This section presents original results from new simulations not reported before. We consider a single spherical particle located at the centre of a simulation box with dimensions $(L_x,L_y,H)=(20R,20R,20R)$. The box's boundary conditions are as in \S\ref{SedimentTransport}: periodic in $x$ and $y$, no-slip at $z=0$, and free-slip at $z=H$. However, there is no longer gravity ($\bm{g}=0$), but the incompressible, Newtonian fluid flow is driven by a constant negative streamwise pressure gradient, $\rho_\mathrm{f}a\equiv-\partial_xP>0$. Physical quantities are now measured in natural units of $\rho_\mathrm{f}$, $a$, and $2R$, that is, the constant acceleration $a$ replaces the acceleration scale $(\rho_\mathrm{s}/\rho_\mathrm{f}-1)|\bm{g}|$ used in \S\ref{SedimentTransport}. In addition to hydrodynamic forces, the particle is subjected to a vertical force that always exactly cancels the vertical hydrodynamic force, in order to keep the vertical location constant at $z=5$ and its vertical centre-of-mass velocity constant at $v_{{\uparrow}z}=0$. Otherwise, for turbulent conditions, the particle would migrate toward the bottom wall, making it much more challenging to obtain unambiguous results. Since the vertical force does not do work, it represents a purely kinematic constraint that does not affect the flow forces acting on the particle, analogous to kinematic constraints in analytical mechanics. It could, for example, be established through giving the particle a charge and imposing an electric field that instantly recognizes and adapts to the hydrodynamics. Note that instant signalling and action are consistent with Newton's physics, which fluid dynamics is based on.

\begin{figure}
\centering
 \includegraphics[width=\textwidth]{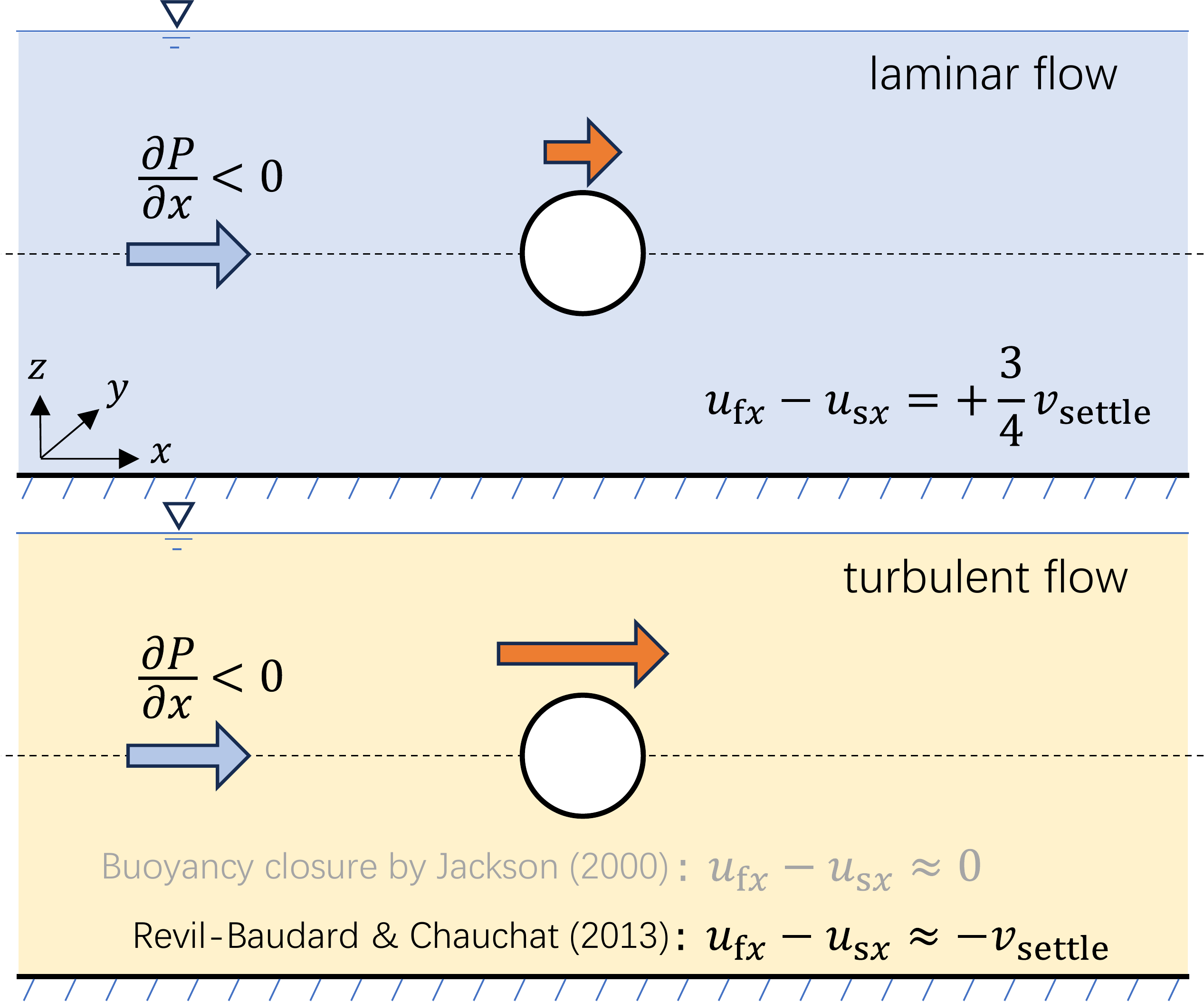}
\caption{Sketch of the horizontal settling scenario.}
\label{HorizontalSettlingSketch}
\end{figure}

The above system has been designed to mimic the vertical settling of a particle in a quiescent fluid, but in the horizontal direction (figure~\ref{HorizontalSettlingSketch}).
While for the vertical settling scenario, the generalized-buoyancy force emerges due to vertical normal stress gradients, for this ``horizontal settling'' scenario, it emerges due to vertical shear stress gradients and the flow-driving horizontal pressure gradient. However, as we argued in \S\ref{BuoyancyCriterion} and as predicted by the closure by \citet{RevilBaudardChauchat13}, Reynolds shear stress gradients, $\d_z\sigma^\mathrm{f}_{\mathrm{Re}zx}$, should not contribute to buoyancy. Therefore, for turbulent flow, where the Reynolds shear stress $\sigma^\mathrm{f}_{\mathrm{Re}zx}$ dominates the effective mechanical shear stress in the average fluid phase momentum balance,
\begin{equation}
 \rho_\mathrm{f}a=-\frac{\d}{\d z}\left(\sigma^\mathrm{f}_{zx}+\sigma^\mathrm{f}_{\mathrm{Re}zx}\right), \label{MomFluidHorizontalSettling}
\end{equation}
we expect that the shear stress gradient force is negligible. Then, in the steady state, the only two significant forces acting on the particle on average are the force from the constant pressure gradient, $(4/3)\pi R^3\rho_\mathrm{f}a$, and the drag force. Hence, the particle should lead the flow by approximately the settling velocity $v_\mathrm{settle}$, which can be calculated by equation~(9) of \citet{Dietrich82a}:
\begin{equation}
\begin{split}
 \log_{10}v_\mathrm{settle}=&-3.76715+1.92944(\log_{10}\mathrm{Ar})-0.09815(\log_{10}\mathrm{Ar})^2 \\
 &-0.00575(\log_{10}\mathrm{Ar})^3+0.00056(\log_{10}\mathrm{Ar})^4, \label{vs}
\end{split}
\end{equation}
where $\mathrm{Ar}\equiv\rho_\mathrm{f}^2a(2R)^3/\eta_\mathrm{f}^2$ (i.e., $\mathrm{Ar}=1/\eta_\mathrm{f}^2$ in natural units) is the Archimedes number. For the averaged velocity fields evaluated at the particle centre location, $z=5$, this means
\begin{equation}
 \text{turbulent flow:}\quad(u_{\mathrm{f}x}-u_{\mathrm{s}x})|_{z=5}\approx-v_\mathrm{settle}. \label{TurbulentExpectation}
\end{equation}
By contrast, if the Reynolds stress actually contributed to buoyancy, as predicted by the Jackson closure, (\ref{BuoyancyJackson}), the shear stress gradient force would compensate the pressure gradient force $(4/3)\pi R^3\rho_\mathrm{f}a$ due to (\ref{MomFluidHorizontalSettling}). Then, only the drag force would act on the particle, and we would therefore expect $u_{\mathrm{f}x}-u_{\mathrm{s}x}\approx0$ at $z=5$. Furthermore, regardless of the chosen buoyancy closure, for fully laminar flow, the shear stress gradient force always exactly compensates the pressure gradient force. For such conditions, not only the bulk flow Reynolds number $\Re$ but also the particle Reynolds numbers $\Re_U$ and $\Re_W$ are small, meaning that the assumptions behind the MRG equation (\ref{MRG}) are approximately satisfied. Then, the counteraction of the shear stress gradient force $F_{\tau x}$ is further enhanced by the Fax\'en correction to the drag force by a value that is equal to $(3/4)F_{\tau x}$. This is because the surface integral $\int_\mathbb{S}(\bm{\tilde u}-\bm{v}_\uparrow)\d S$ in (\ref{MRG}) simplifies to $4\pi R^2(\bm{\tilde u}[\bm{x}_\mathrm{o}]-\bm{v}_\uparrow+R^2\Delta\bm{\tilde u}[\bm{x}_\mathrm{o}]/6)$, since the steady Stokes equations imply $\Delta^k\bm{\tilde u}=0$ for any integer $k>1$ \citep{KimKarrila91}. For fully laminar flow, we therefore expect
\begin{equation}
 \text{laminar flow:}\quad(u_{\mathrm{f}x}-u_{\mathrm{s}x})|_{z=5}=+(3/4)v_\mathrm{settle}. \label{ViscousExpectation}
\end{equation}

The three strikingly different expectations, depending on the flow conditions and chosen buoyancy closure, for the outcome of $(u_{\mathrm{f}x}-u_{\mathrm{s}x})|_{z=5}$ constitute an excellent environment for testing the different closures with our particle-resolving simulations. We simulate both a viscous scenario, $\mathrm{Ar}=1$, and a turbulent scenario, $\mathrm{Ar}=10^4$. The corresponding bulk flow Reynolds numbers are $\Re\approx499$ and $\Re\approx78092$, respectively. To allow a quick adaptation of the particle to the flow velocity, we choose a low particle density, $\rho_\mathrm{s}=0.5$, which is close to the smallest value that does not lead to numerical issues. As in \S\ref{SedimentTransport}, the simulations' grid length $\Delta=0.05$. Once the simulations have reached a steady state, the data are averaged as in (\ref{AveragingDNSDEM}) over the entire periodic bed-tangential domain, $x\in[0,L_x)$ and $y\in[0,L_y)$, and over a sufficient number $N_T$ of timewise \textit{well-separated} instants $t_k$, $N_T=214$ for the viscous and $N_T=2831$ for the turbulent case.

\begin{figure}
\centering
 \includegraphics[width=\textwidth]{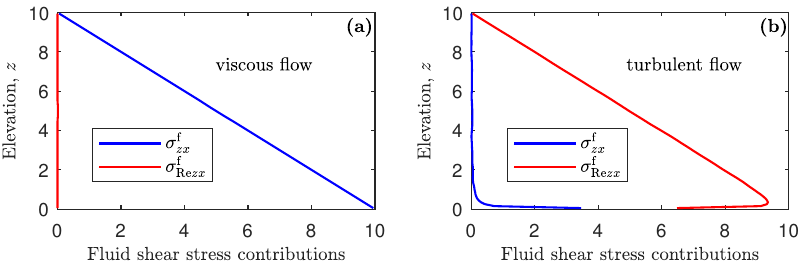}
\caption{Fluid phase shear stress contributions obtained from particle-resolving simulations of the horizontal settling scenario (figure~\ref{HorizontalSettlingSketch}) for (a) viscous flow ($\mathrm{Ar}=1$, $\Re\approx499$) and (b) turbulent flow ($\mathrm{Ar}=10^4$, $\Re\approx78092$).}
\label{SigmaProfileszxSettling}
\end{figure}

\begin{figure}
\centering
 \includegraphics[width=\textwidth]{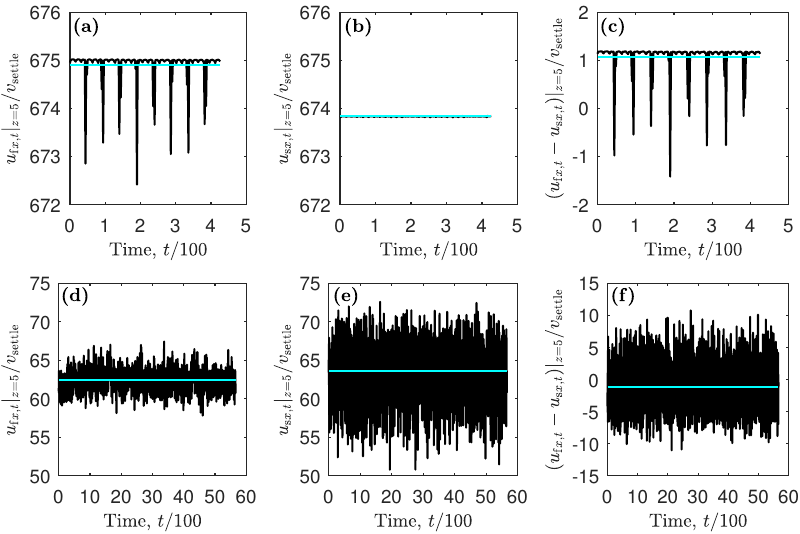}
\caption{Time series of the spatially averaged fluid phase ($u_{\mathrm{f}x,t}$) and dispersed-phase ($u_{\mathrm{s}x,t}$) streamwise flow velocities, and their difference $u_{\mathrm{f}x,t}-u_{\mathrm{s}x,t}$, all evaluated at the particle centre elevation $z=5$. The cyan lines indicate the respective mean values. The data are from particle-resolving simulations of the horizontal settling scenario (figure~\ref{HorizontalSettlingSketch}) for (a)-(c) viscous flow ($\mathrm{Ar}=1$, $\Re\approx499$) and (d)-(f) turbulent flow ($\mathrm{Ar}=10^4$, $\Re\approx78092$).}
\label{TimeSeries}
\end{figure}

\begin{figure}
\centering
 \includegraphics[width=\textwidth]{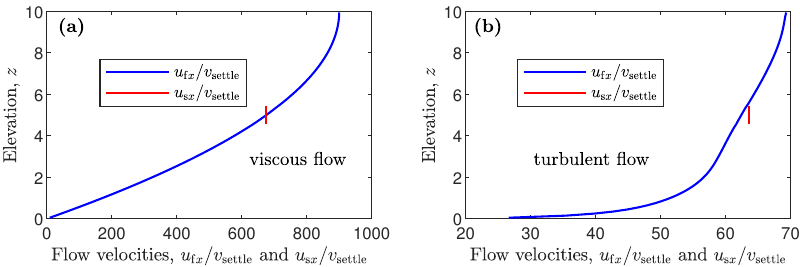}
\caption{Vertical profiles of the fluid-phase-averaged streamwise flow velocity $u_{\mathrm{f}x}$ and the dispersed-phase-averaged streamwise flow velocity $u_{\mathrm{s}x}$, obtained from particle-resolving simulations of the horizontal settling scenario (figure~\ref{HorizontalSettlingSketch}) for (a) viscous flow ($\mathrm{Ar}=1$, $\Re\approx499$) and (b) turbulent flow ($\mathrm{Ar}=10^4$, $\Re\approx78092$).}
\label{VelocityProfilesSettling}
\end{figure}

\begin{figure}
\centering
 \includegraphics[width=\textwidth]{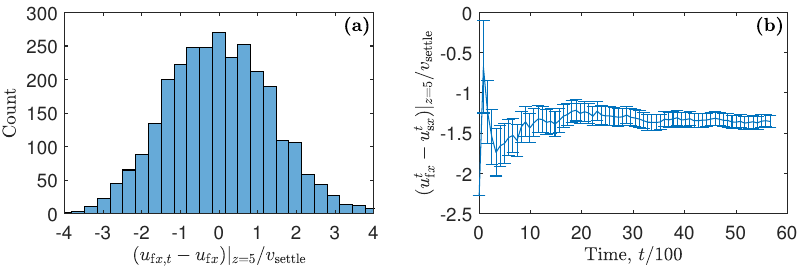}
\caption{(a) Histogram of the spatially averaged fluid phase velocity $u_{\mathrm{f}x,t}$ relative to its mean $u_{\mathrm{f}x}$, evaluated at the particle centre elevation $z=5$. (b) Temporal evolution of the velocity difference averaged up to time $t$, $(u_{\mathrm{f}x}^t-u_{\mathrm{s}x}^t)|_{z=5}$, and its standard error. The data are from particle-resolving simulations of the horizontal settling scenario (figure~\ref{HorizontalSettlingSketch}) for turbulent flow ($\mathrm{Ar}=10^4$, $\Re\approx78092$).}
\label{StatisticalBias}
\end{figure}

Figure~\ref{SigmaProfileszxSettling} confirms that the effective mechanical fluid phase shear stress $\sigma^\mathrm{f}_{zx}$ dominates for the viscous case, whereas the Reynolds shear stress $\sigma^\mathrm{f}_{\mathrm{Re}zx}$ dominates for the turbulent case.
Furthermore, figure~\ref{TimeSeries} shows the time series of the spatially averaged fluid phase ($u_{\mathrm{f}x,t}$) and dispersed-phase ($u_{\mathrm{s}x,t}$) streamwise flow velocities, and their difference $u_{\mathrm{f}x,t}-u_{\mathrm{s}x,t}$, all evaluated at the particle centre elevation $z=5$.
It can be seen that the mean velocity difference $(u_{\mathrm{f}x}-u_{\mathrm{s}x})|_{z=5}$ (cyan lines in figures~\ref{TimeSeries}(c),(f)) is positive ($\approx1.1v_\mathrm{settle}$) in the viscous case and negative ($\approx-1.2v_\mathrm{settle}$) in the turbulent case. Since these values are not far off the predictions $(3/4)v_\mathrm{settle}$ and $-v_\mathrm{settle}$ by (\ref{ViscousExpectation}) and (\ref{TurbulentExpectation}), respectively, they strongly support the buoyancy closure by \citet{RevilBaudardChauchat13}, (\ref{BuoyancyChauchat}), over the Jackson closure, (\ref{BuoyancyJackson}). In addition, the vertical profiles of $u_{\mathrm{f}x}$ and $u_{\mathrm{s}x}$ reveal that $u_{\mathrm{f}x}$ remains smaller than $u_{\mathrm{s}x}$ for all locations $z\in(4.5,5.5)$ covered by the particle in the turbulent case (figure~\ref{VelocityProfilesSettling}(b)).
This would not be physically possible if a Reynolds shear stress gradient force compensated the pressure gradient force. Finally, figure~\ref{StatisticalBias} confirms that the findings for the turbulent case are neither the result of correlations between the turbulent fluid velocity fluctuations and drag forces---this possibility is ruled out by the fact that the distribution of $u_{\mathrm{f}x,t}|_{z=5}$ is nearly symmetric relative to its mean $u_{\mathrm{f}x}|_{z=5}$ (figure~\ref{StatisticalBias}(a))---nor caused by insufficient averaging (figure~\ref{StatisticalBias}(b)).
Put together, figures~\ref{TimeSeries}-\ref{StatisticalBias} falsify the Jackson closure, (\ref{BuoyancyJackson}).

\subsection{Thought experiment: dilute Stokesian suspension} \label{StressFluidParticleInteractions}
We consider a mixture of an incompressible Newtonian fluid and identical spherical particles at vanishingly small particle Reynolds numbers and Stokes number, $\Re_U\sim0$, $\Re_W\sim0$, and $\mathrm{St}\equiv\rho_\mathrm{s}R^2/(T\eta_\mathrm{f})\sim0$, where $T$ is the characteristic time over which the disturbance flow velocity changes. It means that transients in the Stokes disturbance flow for any given particle can also be neglected. Furthermore, the dispersed-phase volume fraction $\beta_\mathrm{s}$ shall be non-zero but sufficiently low to ensure that fluid-mediated particle-particle interactions are negligible, while the sphere radius $R$ shall be sufficiently smaller than the characteristic length $L$ over which the macroscopic flow changes. Under these assumptions, the following average fluid phase momentum balance can be derived \citep[equation~(2.46) of][]{Jackson00}:
\begin{equation}
\begin{split}
 \rho_\mathrm{f}\D^\mathrm{f}_t\bm{u}_\mathrm{f}&=\bm{\nabla}\cdot\left\{-\langle P\rangle^\mathrm{f}\bm{1}+\eta_\mathrm{f}\left(\bm{\nabla}\langle\bm{u}\rangle+(\bm{\nabla}\langle\bm{u}\rangle)^\mathrm{T}\right)+\frac{5\eta_\mathrm{f}\beta_\mathrm{s}}{2}\left(\bm{\nabla}\bm{u}_\mathrm{f}+\left(\bm{\nabla}\bm{u}_\mathrm{f}\right)^\mathrm{T}\right)\right. \\
 &+\left.3\eta_\mathrm{f}\beta_\mathrm{s}\left(\frac{1}{2}\left(\bm{\nabla}\bm{u}_\mathrm{f}-\left(\bm{\nabla}\bm{u}_\mathrm{f}\right)^\mathrm{T}\right)-\bm{\epsilon}\cdot\langle\bm{\omega}\rangle^\mathrm{s}\right)-\bm{\nabla}\left[\frac{3\eta_\mathrm{f}\beta_\mathrm{s}}{4}\left(\bm{u}_\mathrm{f}-\bm{u}_\mathrm{s}\right)\right]\right\} \\
 &-\frac{9\eta_\mathrm{f}\beta_\mathrm{s}}{2R^2}\left(\bm{u}_\mathrm{f}-\bm{u}_\mathrm{s}+\frac{R^2}{6}\Delta\bm{u}_\mathrm{f}\right)+\rho_\mathrm{f}\bm{g},
\end{split} \label{MomJackson}
\end{equation}
where $\Delta=\bm{\nabla}\cdot\bm{\nabla}$ is the Laplace operator, $\bm{\epsilon}$ the total antisymmetric tensor, $P$ the instantaneous, local fluid pressure, and $\bm{\omega}\equiv\sum_pX^p\bm{\omega}^p$ the particle-angular velocity field. Note that, for the considered scenario, $\bm{\sigma^\mathrm{f}_\mathrm{Re}}=0$ due to $\Re_W\sim0$ and $\mathrm{St}\sim0$ \citep{Jackson97} and $\langle\bm{u}\rangle=\beta_\mathrm{f}\bm{u}_\mathrm{f}+\beta_\mathrm{s}\bm{u}_\mathrm{s}$ due to constant microscopic fluid and particle densities.

The question is now how to interpret the different terms in (\ref{MomJackson}). Due to $\Re_U\sim0$, $\Re_W\sim0$, and $\mathrm{St}\sim0$, the disturbance flow is described by the steady Stokes equations. Hence, the instantaneous disturbance or generalized-drag force acting on any given particle centred at $\bm{x}_\mathrm{o}$ is described by only the drag force and its Fax\'en correction in the MRG equation (\ref{MRG}). In particular, the surface integral $\int_\mathbb{S}(\bm{\tilde u}-\bm{v}_\uparrow)\d S$ in (\ref{MRG}) simplifies to $4\pi R^2(\bm{\tilde u}[\bm{x}_\mathrm{o}]-\bm{v}_\uparrow+R^2\Delta\bm{\tilde u}[\bm{x}_\mathrm{o}]/6)$, since the steady Stokes equations imply $\Delta^k\bm{\tilde u}=0$ for any integer $k>1$ \citep{KimKarrila91}. Therefore, identifying $\bm{u}_\mathrm{f}$ and $\bm{u}_\mathrm{s}$ as the analogues to $\bm{\tilde u}[\bm{x}_\mathrm{o}]$ and $\bm{v}_\uparrow$ in (\ref{MRG}), respectively, the first term on the third line of (\ref{MomJackson}) is recognized as the negative sum of the average drag force and its Fax\'en correction per unit volume of the pure fluid. However, the actual generalized-drag force density $\beta_\mathrm{s}\langle\bm{f}_\mathrm{D}\rangle^\mathrm{s}$ is the sum of the average drag force and its Fax\'en correction per unit volume of the fluid-particle mixture, not of the pure fluid. This is because fluid phase momentum balances in two-fluid models refer to the mixture volume. To translate pure-fluid-based volumetric densities into mixture-based volumetric densities, one has to multiply the former with the fluid volume fraction $\beta_\mathrm{f}$. Hence, we conclude
\begin{equation}
 \beta_\mathrm{s}\langle\bm{f}_\mathrm{D}\rangle^\mathrm{s}=\frac{9\eta_\mathrm{f}\beta_\mathrm{s}\beta_\mathrm{f}}{2R^2}\left(\bm{u}_\mathrm{f}-\bm{u}_\mathrm{s}+\frac{R^2}{6}\Delta\bm{u}_\mathrm{f}\right). \label{DragJackson97}
\end{equation}
The remaining part $(\beta_\mathrm{s}^2/\beta_\mathrm{f})\langle\bm{f}_\mathrm{D}\rangle^\mathrm{s}$ is therefore attributed to the generalized-buoyancy force density $\beta_\mathrm{s}\langle\bm{f}_\mathrm{B}\rangle^\mathrm{s}$ alongside the usual gravitational and inertial terms:
\begin{equation}
 \beta_\mathrm{s}\langle\bm{f}_\mathrm{B}\rangle^\mathrm{s}=\beta_\mathrm{s}\left(\D^\mathrm{f}_t\bm{u}_\mathrm{f}-\rho_\mathrm{f}\bm{g}+\beta_\mathrm{f}^{-1}\beta_\mathrm{s}\langle\bm{f}_\mathrm{D}\rangle^\mathrm{s}\right). \label{BuoyancyJackson97}
\end{equation}
This somewhat odd appearance of drag terms in $\beta_\mathrm{s}\langle\bm{f}_\mathrm{B}\rangle^\mathrm{s}$ was exactly what led to the debate between \citet{Cliftetal87} and others that we briefly discussed in \S\ref{Introduction}.

The remaining terms in (\ref{MomJackson}) are the gravitational body force density $\beta_\mathrm{f}\rho_\mathrm{f}\bm{g}$ and a divergence of two distinct stress tensors: the fluid-phase-averaged stress tensor \citep{Jackson97},
\begin{equation}
 \beta_\mathrm{f}\langle\bm{\sigma}\rangle^\mathrm{f}=-\beta_\mathrm{f}\langle P\rangle^\mathrm{f}\bm{1}+\eta_\mathrm{f}\left(\bm{\nabla}\langle\bm{u}\rangle+(\bm{\nabla}\langle\bm{u}\rangle)^\mathrm{T}\right), \label{AverageStress}
\end{equation}
and a pseudo-stress tensor $\bm{\sigma^\mathrm{f}_\mathrm{s}}$ that emerges due to fluid-particle interactions \citep{Jackson97},
\begin{equation}
\begin{split}
 \bm{\sigma^\mathrm{f}_\mathrm{s}}=&-\beta_\mathrm{s}\langle P\rangle^\mathrm{f}\bm{1}+\frac{5\eta_\mathrm{f}\beta_\mathrm{s}}{2}\left(\bm{\nabla}\bm{u}_\mathrm{f}+\left(\bm{\nabla}\bm{u}_\mathrm{f}\right)^\mathrm{T}\right) \\
 &+3\eta_\mathrm{f}\beta_\mathrm{s}\left(\frac{1}{2}\left(\bm{\nabla}\bm{u}_\mathrm{f}-\left(\bm{\nabla}\bm{u}_\mathrm{f}\right)^\mathrm{T}\right)-\bm{\epsilon}\cdot\langle\bm{\omega}\rangle^\mathrm{s}\right)-\bm{\nabla}\left[\frac{3\eta_\mathrm{f}\beta_\mathrm{s}}{4}\left(\bm{u}_\mathrm{f}-\bm{u}_\mathrm{s}\right)\right].
\end{split} \label{PseudoStressJackson}
\end{equation}
Using (\ref{DragJackson97})-(\ref{PseudoStressJackson}) and $\bm{\sigma^\mathrm{f}}\equiv\beta_\mathrm{f}\langle\bm{\sigma}\rangle^\mathrm{f}+\bm{\sigma^\mathrm{f}_\mathrm{s}}$, we can rewrite (\ref{MomJackson}) in either of the following two forms:
\begin{align}
 \beta_\mathrm{f}\rho_\mathrm{f}\D^\mathrm{f}_t\bm{u}_\mathrm{f}&=\bm{\nabla}\cdot\bm{\sigma^\mathrm{f}}+\beta_\mathrm{f}\rho_\mathrm{f}\bm{g}-\beta_\mathrm{s}\langle\bm{f}_\mathrm{D}\rangle^\mathrm{s}-\beta_\mathrm{s}\langle\bm{f}_\mathrm{B}\rangle^\mathrm{s}, \label{MomJackson2} \\
 \rho_\mathrm{f}\D^\mathrm{f}_t\bm{u}_\mathrm{f}&=\bm{\nabla}\cdot\bm{\sigma^\mathrm{f}}+\rho_\mathrm{f}\bm{g}-\beta_\mathrm{f}^{-1}\beta_\mathrm{s}\langle\bm{f}_\mathrm{D}\rangle^\mathrm{s}. \label{MomJackson3}
\end{align}
Multiplying (\ref{MomJackson3}) with $\beta_\mathrm{f}$ and subtracting the result from (\ref{MomJackson2}) then yields the buoyancy closure by \citet{RevilBaudardChauchat13}, (\ref{BuoyancyChauchat}), here equivalent to the Jackson closure, (\ref{BuoyancyJackson}), as $\bm{\sigma^\mathrm{f}_\mathrm{Re}}=0$ for the considered scenario.

\subsection{Conclusion from the particle-resolving simulations and thought experiment} \label{ConclusionPseudoStresses}
The results presented in \S\ref{Simulations} and \S\ref{StressFluidParticleInteractions} strongly support the buoyancy closure by \citet{RevilBaudardChauchat13}, (\ref{BuoyancyChauchat}), over the closure by \citet{Zhangetal07b}, (\ref{BuoyancyZhang}), and the Jackson closure, (\ref{BuoyancyJackson}). However, as we argued in \S\ref{BuoyancyCriterion}, (\ref{BuoyancyChauchat}) constitutes only the small-particle approximation of what we consider to be the physically appropriate closure. As such, it leads to an inconsistency for some fluid-particle systems at complete rest (\S\ref{SmallParticleApproximation}). It is therefore generalized to arbitrarily large particles in the next section.

\section{Unique consistent buoyancy closure} \label{NewClosure}
In this section, we derive the unique buoyancy closure (\ref{BuoyancyHypothesis}) that satisfies the criterion (\ref{Criterion}) and analyse some of its consequences and mathematical properties. The derivation makes use of operator analysis, which requires introducing new notation first.

\subsection{Operator notation} \label{OperatorNotation}
The operators discussed in this section map an arbitrary macroscopic tensor fields $\bm{A}$ (i.e., fields that satisfy $\langle\bm{A}\rangle=\bm{A}$) in space and time, $\mathbb{R}^3\times\mathbb{R}$, to new macroscopic tensor fields in $\mathbb{R}^3\times\mathbb{R}$ of the same rank. Their actions are the same for each of the $1$ (scalar), $3$ (vector), or $9$ (rank-$2$ tensor) components of $\bm{A}$.

We define the operator $\mathcal{L}$ through the action
\begin{equation}
 \mathcal{L}\beta_\mathrm{s}\bm{A}\equiv\left\langle\sum\nolimits_pX^p\underline{\bm{A}}_{\mathbb{V}^p}\right\rangle=\left\langle X_\mathrm{s}\sum\nolimits_pX^p\underline{\bm{A}}_{\mathbb{V}^p}\right\rangle=\beta_\mathrm{s}\left\langle\sum\nolimits_pX^p\underline{\bm{A}}_{\mathbb{V}^p}\right\rangle^\mathrm{s}, \label{L}
\end{equation}
where we introduced the notation $\underline{\cdot}_{\mathbb{V}^p}\equiv\frac{1}{V^p}\int_{\mathbb{V}^p}\cdot\d V$ and used $X_\mathrm{s}X^p=X^p$ and (\ref{PhaseAverages}). Equation~(\ref{L}) means that $\mathcal{L}$ first averages $\bm{A}$ over a given particle $p$'s domain $\mathbb{V}^p$. Then, it assigns the resulting average to each point of the same domain, $\bm{x}\in\mathbb{V}^p$, but to nowhere else. It does so for all particles $p$, superposes all these new fields, and macroscopically averages this superposition. The space-time dependence of the resulting field emerges from the generalized functions $X^p[\bm{x},t]$ within the averaging procedure $\langle\cdot\rangle$.

The operator $\mathcal{L}$ and the identity operator $\mathcal{I}$ appear in the identity \citep[Supplementary material of][]{Pahtzetal26}
\begin{equation}
 \bm{\nabla}\cdot[\bm{A}]_\mathrm{s}^\prime=(\mathcal{L}-\mathcal{I})\beta_\mathrm{s}\bm{\nabla}\cdot\bm{A}, \label{FluctuationOperator1}
\end{equation}
where $[\cdot]_\mathrm{s}^\prime$ is another operator. Its detailed micromechanical action is given by (\ref{[]}), but it is not relevant for what comes later. It is only relevant that $[\cdot]_\mathrm{s}^\prime$ exists. Note that, in the small-particle approximation, $[\cdot]_\mathrm{s}^\prime\approx0$ and $\mathcal{L}\approx\mathcal{I}$.

For the coming analysis, we introduce the convention that operators written using capital letters, such as $\mathcal{L}$ and $\mathcal{I}$, shall act on all terms right of them. The exception is when \textit{both} capital-letter operator and operated field are surrounded by parentheses; for example, $(\mathcal{L}\beta_\mathrm{s})\bm{A}=\beta_\mathrm{s}\bm{A}$ due to $\mathcal{L}\beta_\mathrm{s}=\beta_\mathrm{s}$, whereas $(\mathcal{L}-\mathcal{I})\beta_\mathrm{s}\bm{A}$ should be read as $\mathcal{L}\beta_\mathrm{s}\bm{A}-\beta_\mathrm{s}\bm{A}$. Furthermore, we introduce the placeholder symbol ``$\underline{~~}$'', which shall indicate that the operator action is not done even when surrounded by parentheses; for example, $(\mathcal{L}\beta_\mathrm{s}\underline{~~})\bm{A}=\mathcal{L}\beta_\mathrm{s}\bm{A}$. Objects such as $\mathcal{L}\beta_\mathrm{s}\underline{~~}$ can therefore be interpreted as operators themselves. Moreover, when the operation $\cdot^{-1}$ is applied on an operator $\mathcal{A}$, it shall refer to its inverse, defined through the property $\mathcal{A}^{-1}\mathcal{A}=\mathcal{A}\mathcal{A}^{-1}=\mathcal{I}$, while $\mathcal{A}^k$ and $[\cdot]_\mathrm{s}^{\prime k}$ shall mean $k$ applications of $\mathcal{A}$ and $[\cdot]_\mathrm{s}^\prime$, respectively.

\subsection{Derivation of closure} \label{Derivation}
The definition (\ref{L}) of the operator $\mathcal{L}$ allow us to rewrite (\ref{BuoyancyHypothesis}) as
\begin{equation}
 \beta_\mathrm{s}\langle\bm{f}_\mathrm{B}\rangle^\mathrm{s}=\beta_\mathrm{s}\left\langle\sum\nolimits_p\frac{X^p}{V^p}\int_{\mathbb{S}^p}\bm{n}^p\cdot\bm{\hat\sigma^\mathrm{f}}\d S\right\rangle^\mathrm{s}=\left\langle\sum\nolimits_p\frac{X^p}{V^p}\int_{\mathbb{V}^p}\bm{\nabla}\cdot\bm{\hat\sigma^\mathrm{f}}\d V\right\rangle=\mathcal{L}\beta_\mathrm{s}\bm{\nabla}\cdot\bm{\hat\sigma^\mathrm{f}}, \label{BuoyancyNew0}
\end{equation}
where we used $X_\mathrm{s}X^p=X^p$ and (\ref{PhaseAverages}). Then, we can restate the buoyancy criterion (\ref{Criterion}) in three different manners as
\begin{align}
 \bm{\nabla}\cdot\bm{\sigma^\mathrm{f}}&=\beta_\mathrm{f}\bm{\nabla}\cdot\bm{\hat\sigma^\mathrm{f}}+\mathcal{L}\beta_\mathrm{s}\bm{\nabla}\cdot\bm{\hat\sigma^\mathrm{f}}, \label{Criterion1} \\
 \bm{\nabla}\cdot\bm{\sigma^\mathrm{f}}&=\bm{\nabla}\cdot\left(\bm{\hat\sigma^\mathrm{f}}+\left[\bm{\hat\sigma^\mathrm{f}}\right]_\mathrm{s}^\prime\right), \label{Criterion2} \\
 \mathcal{L}\beta_\mathrm{s}\beta_\mathrm{f}^{-1}\bm{\nabla}\cdot\bm{\sigma^\mathrm{f}}&=\beta_\mathrm{s}\langle\bm{f}_\mathrm{B}\rangle^\mathrm{s}+\mathcal{L}\beta_\mathrm{s}\beta_\mathrm{f}^{-1}\beta_\mathrm{s}\langle\bm{f}_\mathrm{B}\rangle^\mathrm{s}, \label{Criterion1_2}
\end{align}
where (\ref{Criterion2}) uses (\ref{FluctuationOperator1}) and (\ref{Criterion1_2}) follows from applying $\mathcal{L}\beta_\mathrm{s}\beta_\mathrm{f}^{-1}\underline{~~}$ on (\ref{Criterion1}) using (\ref{BuoyancyNew0}). Inverting these relations yields
\begin{align}
 \bm{\nabla}\cdot\bm{\hat\sigma^\mathrm{f}}&=\left(\mathcal{L}\beta_\mathrm{s}\underline{~~}+\mathcal{I}\beta_\mathrm{f}\underline{~~}\right)^{-1}\bm{\nabla}\cdot\bm{\sigma^\mathrm{f}}=\sum_{k=0}^\infty(-1)^k\left(\mathcal{L}\beta_\mathrm{s}\underline{~~}-\mathcal{I}\beta_\mathrm{s}\underline{~~}\right)^k\bm{\nabla}\cdot\bm{\sigma^\mathrm{f}}, \label{CriterionInverted1} \\
 \bm{\hat\sigma^\mathrm{f}}&=\left([\cdot]^\prime_\mathrm{s}+\mathcal{I}\right)^{-1}\bm{\sigma^\mathrm{f}}=\sum_{k=0}^\infty(-1)^k\left[\bm{\sigma^\mathrm{f}}\right]_\mathrm{s}^{\prime k}, \label{CriterionInverted2} \\
 \beta_\mathrm{s}\langle\bm{f}_\mathrm{B}\rangle^\mathrm{s}&=\left(\mathcal{L}\beta_\mathrm{s}\beta_\mathrm{f}^{-1}\underline{~~}+\mathcal{I}\right)^{-1}\mathcal{L}\beta_\mathrm{s}\beta_\mathrm{f}^{-1}\bm{\nabla}\cdot\bm{\sigma^\mathrm{f}}=\sum_{k=0}^\infty(-1)^k\left(\mathcal{L}\beta_\mathrm{s}\beta_\mathrm{f}^{-1}\underline{~~}\right)^{k+1}\bm{\nabla}\cdot\bm{\sigma^\mathrm{f}}, \label{BuoyancyNew1}
\end{align}
respectively, where the right-hand sides exploit the Neumann series representation of operators. Convergence of these series is often nearly guaranteed in (\ref{CriterionInverted1}) and (\ref{CriterionInverted2}) as $\beta_\mathrm{s}$ is usually substantially smaller than unity and $[\bm{\sigma^\mathrm{f}}]_\mathrm{s}^{\prime k}\sim\beta_\mathrm{s}^k$, while $\mathcal{L}-\mathcal{I}$ and $[\cdot]^\prime_\mathrm{s}$ typically do not increase the magnitude of a field. However, (\ref{BuoyancyNew1}) requires at least $\beta_\mathrm{s}\beta_\mathrm{f}^{-1}<1$ for convergence, i.e., $\beta_\mathrm{s}<0.5$. Equation~(\ref{CriterionInverted2}) means that a stress $\bm{\hat\sigma^\mathrm{f}}$ that satisfies the buoyancy criterion (\ref{Criterion}) does, indeed, exist, justifying the physical rationale laid out in \S\ref{BuoyancyCriterion}. In particular, for fluid-particle systems at complete rest, where $\bm{\sigma^\mathrm{f}}=\langle\bm{\sigma}\rangle^\mathrm{f}+[\langle\bm{\sigma}\rangle^\mathrm{f}]^\prime_\mathrm{s}$ due to (\ref{SigmaFluid}) and $\bm{\sigma^\mathrm{f}_\mathrm{s^\prime}}=0$, it reduces to $ \bm{\hat\sigma^\mathrm{f}}=\langle\bm{\sigma}\rangle^\mathrm{f}$, as required. In addition, (\ref{BuoyancyNew1}) constitutes a first explicit expression (but with a not necessarily convergent series expansion) linking $\beta_\mathrm{s}\langle\bm{f}_\mathrm{B}\rangle^\mathrm{s}$ to the effective mechanical stress $\bm{\sigma^\mathrm{f}}$. Another equivalent explicit expression (with an almost always convergent series expansion) results from applying $\mathcal{L}\beta_\mathrm{s}\underline{~~}$ on (\ref{CriterionInverted1}) using (\ref{BuoyancyNew0}):
\begin{equation}
 \beta_\mathrm{s}\langle\bm{f}_\mathrm{B}\rangle^\mathrm{s}=\mathcal{L}\beta_\mathrm{s}\left(\mathcal{L}\beta_\mathrm{s}\underline{~~}+\mathcal{I}\beta_\mathrm{f}\underline{~~}\right)^{-1}\bm{\nabla}\cdot\bm{\sigma^\mathrm{f}}. \label{BuoyancyNew1_2}
\end{equation}
Yet another equivalent explicit expression is obtained from rearranging (\ref{MomFluidHypothesis}) for $\bm{\nabla}\cdot\bm{\hat\sigma^\mathrm{f}}$ and substituting the result in (\ref{BuoyancyNew0}):
\begin{equation}
 \beta_\mathrm{s}\langle\bm{f}_\mathrm{B}\rangle^\mathrm{s}=\mathcal{L}\beta_\mathrm{s}\left(\rho_\mathrm{f}\D^\mathrm{f}_t\bm{u}_\mathrm{f}-\rho_\mathrm{f}\bm{g}-\beta_\mathrm{f}^{-1}\bm{\nabla}\cdot\bm{\sigma^\mathrm{f}_\mathrm{Re}}+\beta_\mathrm{f}^{-1}\beta_\mathrm{s}\langle\bm{f}_\mathrm{D}\rangle^\mathrm{s}\right). \label{BuoyancyNew2}
\end{equation}
This form is particularly useful for practical applications, since it involves the operator $\mathcal{L}$ only once and not in its inverted form. Yet another equivalent buoyancy expression follows from combining (\ref{Criterion}) and (\ref{Criterion1}):
\begin{equation}
 \beta_\mathrm{s}\langle\bm{f}_\mathrm{B}\rangle^\mathrm{s}=\beta_\mathrm{s}\bm{\nabla}\cdot\bm{\hat\sigma^\mathrm{f}}+\bm{\nabla}\cdot\left[\bm{\hat\sigma^\mathrm{f}}\right]_\mathrm{s}^\prime. \label{BuoyancyNew3}
\end{equation}

\subsection{Resulting two-fluid momentum balances}
Substituting the three equivalent expressions (\ref{BuoyancyNew1_2})-(\ref{BuoyancyNew3}) in (\ref{MomFluid}) and (\ref{MomSolid}) yields three equivalent versions of the average two-fluid momentum balances. Equations~(\ref{CriterionInverted1}) and (\ref{BuoyancyNew1_2}) result in
\begin{align}
 \beta_\mathrm{f}\rho_\mathrm{f}\D^\mathrm{f}_t\bm{u}_\mathrm{f}&=\beta_\mathrm{f}\left(\mathcal{L}\beta_\mathrm{s}\underline{~~}+\mathcal{I}\beta_\mathrm{f}\underline{~~}\right)^{-1}\bm{\nabla}\cdot\bm{\sigma^\mathrm{f}}+\bm{\nabla}\cdot\bm{\sigma^\mathrm{f}_\mathrm{Re}}+\beta_\mathrm{f}\rho_\mathrm{f}\bm{g}-\beta_\mathrm{s}\langle\bm{f}_\mathrm{D}\rangle^\mathrm{s}, \label{MomFluid1} \\
 \beta_\mathrm{s}\rho_\mathrm{s}\D^\mathrm{s}_t\bm{u}_\mathrm{s}&=\mathcal{L}\beta_\mathrm{s}\left(\mathcal{L}\beta_\mathrm{s}\underline{~~}+\mathcal{I}\beta_\mathrm{f}\underline{~~}\right)^{-1}\bm{\nabla}\cdot\bm{\sigma^\mathrm{f}}+\bm{\nabla}\cdot\bm{\sigma^\mathrm{s}}+\bm{\nabla}\cdot\bm{\sigma^\mathrm{s}_\mathrm{Re}}+\beta_\mathrm{s}\rho_\mathrm{s}\bm{g}+\beta_\mathrm{s}\langle\bm{f}_\mathrm{D}\rangle^\mathrm{s}. \label{MomSolid1}
\end{align}
Equation~(\ref{BuoyancyNew2}) leads to
\begin{align}
\begin{split}
 \beta_\mathrm{f}\rho_\mathrm{f}\D^\mathrm{f}_t\bm{u}_\mathrm{f}&=\bm{\nabla}\cdot\bm{\sigma^\mathrm{f}}+\bm{\nabla}\cdot\bm{\sigma^\mathrm{f}_\mathrm{Re}}+\beta_\mathrm{f}\rho_\mathrm{f}\bm{g}-\beta_\mathrm{s}\langle\bm{f}_\mathrm{D}\rangle^\mathrm{s} \\
 &-\mathcal{L}\beta_\mathrm{s}\left(\rho_\mathrm{f}\D^\mathrm{f}_t\bm{u}_\mathrm{f}-\rho_\mathrm{f}\bm{g}-\beta_\mathrm{f}^{-1}\bm{\nabla}\cdot\bm{\sigma^\mathrm{f}_\mathrm{Re}}+\beta_\mathrm{f}^{-1}\beta_\mathrm{s}\langle\bm{f}_\mathrm{D}\rangle^\mathrm{s}\right),
\end{split} \label{MomFluid2} \\
\begin{split}
 \beta_\mathrm{s}\rho_\mathrm{s}\D^\mathrm{s}_t\bm{u}_\mathrm{s}&=\bm{\nabla}\cdot\bm{\sigma^\mathrm{s}}+\bm{\nabla}\cdot\bm{\sigma^\mathrm{s}_\mathrm{Re}}+\beta_\mathrm{s}\rho_\mathrm{s}\bm{g}+\beta_\mathrm{s}\langle\bm{f}_\mathrm{D}\rangle^\mathrm{s} \\
 &+\mathcal{L}\beta_\mathrm{s}\left(\rho_\mathrm{f}\D^\mathrm{f}_t\bm{u}_\mathrm{f}-\rho_\mathrm{f}\bm{g}-\beta_\mathrm{f}^{-1}\bm{\nabla}\cdot\bm{\sigma^\mathrm{f}_\mathrm{Re}}+\beta_\mathrm{f}^{-1}\beta_\mathrm{s}\langle\bm{f}_\mathrm{D}\rangle^\mathrm{s}\right).
\end{split} \label{MomSolid2}
\end{align}
Equations~(\ref{BuoyancyNew3}) yields
\begin{align}
 \beta_\mathrm{f}\rho_\mathrm{f}\D^\mathrm{f}_t\bm{u}_\mathrm{f}&=\beta_\mathrm{f}\bm{\nabla}\cdot\bm{\hat\sigma^\mathrm{f}}+\bm{\nabla}\cdot\bm{\sigma^\mathrm{f}_\mathrm{Re}}+\beta_\mathrm{f}\rho_\mathrm{f}\bm{g}-\beta_\mathrm{s}\langle\bm{f}_\mathrm{D}\rangle^\mathrm{s}, \label{MomFluid3} \\
 \beta_\mathrm{s}\rho_\mathrm{s}\D^\mathrm{s}_t\bm{u}_\mathrm{s}&=\beta_\mathrm{s}\bm{\nabla}\cdot\bm{\hat\sigma^\mathrm{f}}+\bm{\nabla}\cdot\bm{\hat\sigma^\mathrm{s}}+\bm{\nabla}\cdot\bm{\sigma^\mathrm{s}_\mathrm{Re}}+\beta_\mathrm{s}\rho_\mathrm{s}\bm{g}+\beta_\mathrm{s}\langle\bm{f}_\mathrm{D}\rangle^\mathrm{s}, \label{MomSolid3}
\end{align}
where (\ref{MomFluid3}) is the same as (\ref{MomFluidHypothesis}) and $\bm{\hat\sigma^\mathrm{s}}\equiv\bm{\sigma^\mathrm{s}}+[\bm{\hat\sigma^\mathrm{f}}]_\mathrm{s}^\prime$ is a modified effective mechanical dispersed-phase stress tensor. Furthermore, (\ref{MomFluid3}) and (\ref{MomSolid3}) can be rearranged and brought into the form
\begin{align}
\begin{split}
 \beta_\mathrm{f}\rho_\mathrm{f}\D^\mathrm{f}_t\bm{u}_\mathrm{f}&=\bm{\nabla}\cdot\bm{\hat\sigma^\mathrm{f}}+\bm{\nabla}\cdot\bm{\sigma^\mathrm{f}_\mathrm{Re}}+\beta_\mathrm{f}\rho_\mathrm{f}\bm{g}-\beta_\mathrm{s}\langle\bm{f}_\mathrm{D}\rangle^\mathrm{s} \\
 &-\beta_\mathrm{s}\left(\rho_\mathrm{f}\D^\mathrm{f}_t\bm{u}_\mathrm{f}-\rho_\mathrm{f}\bm{g}-\beta_\mathrm{f}^{-1}\bm{\nabla}\cdot\bm{\sigma^\mathrm{f}_\mathrm{Re}}+\beta_\mathrm{f}^{-1}\beta_\mathrm{s}\langle\bm{f}_\mathrm{D}\rangle^\mathrm{s}\right),
\end{split} \label{MomFluid4} \\
\begin{split}
 \beta_\mathrm{s}\rho_\mathrm{s}\D^\mathrm{s}_t\bm{u}_\mathrm{s}&=\bm{\nabla}\cdot\bm{\hat\sigma^\mathrm{s}}+\bm{\nabla}\cdot\bm{\sigma^\mathrm{s}_\mathrm{Re}}+\beta_\mathrm{s}\rho_\mathrm{s}\bm{g}+\beta_\mathrm{s}\langle\bm{f}_\mathrm{D}\rangle^\mathrm{s} \\
 &+\beta_\mathrm{s}\left(\rho_\mathrm{f}\D^\mathrm{f}_t\bm{u}_\mathrm{f}-\rho_\mathrm{f}\bm{g}-\beta_\mathrm{f}^{-1}\bm{\nabla}\cdot\bm{\sigma^\mathrm{f}_\mathrm{Re}}+\beta_\mathrm{f}^{-1}\beta_\mathrm{s}\langle\bm{f}_\mathrm{D}\rangle^\mathrm{s}\right).
\end{split} \label{MomSolid4}
\end{align}
The former two versions of the average two-fluid momentum balances, (\ref{MomFluid1})-(\ref{MomSolid2}), are based on the effective mechanical stress tensors $\bm{\sigma^\mathrm{f}}$ and $\bm{\sigma^\mathrm{s}}$. They are of a mathematical structure never reported before due to the explicit appearance of the operator $\mathcal{L}$. By contrast, the latter two versions, (\ref{MomFluid3})-(\ref{MomSolid4}), are of the same mathematical structure as those reported by \citet{RevilBaudardChauchat13}, since the effects of $\mathcal{L}$ are buried in the modified effective mechanical stress tensors $\bm{\hat\sigma^\mathrm{f}}$ and $\bm{\hat\sigma^\mathrm{s}}$. Note that $\bm{\hat\sigma^\mathrm{s}}$ does not only differ quantitatively but also qualitatively from $\bm{\sigma^\mathrm{s}}$, since, in contrast to $\bm{\sigma^\mathrm{s}}$, it contains a stress contribution, $[\bm{\hat\sigma^\mathrm{f}}]_\mathrm{s}^\prime$, that originated in the fluid phase momentum balance.

\subsection{Effect of $\mathcal{L}$ for a mixture of continuous fluid and identically-sized spheres}
\begin{figure}
\centering
 \includegraphics[width=0.5\textwidth]{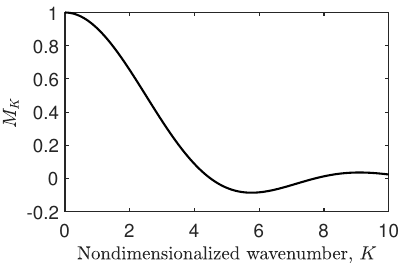}
\caption{The Fourier transform $M_K$ of the volume averaging operator $\mathcal{M}$ essentially acts like a low-pass filter, cutting off contributions from wavenumber vectors $\bm{k}$ with $K\equiv R|\bm{k}|\gtrsim4$.}
\label{Fig_Mhat}
\end{figure}

The mathematical effect of the operator $\mathcal{L}$ can be fully understood for the simplified case of a disperse two-phase flow in which all particles are spheres of the same radius $R$ and volume $V=4\pi R^3/3$. In this case, the action of $\mathcal{L}$ can be expressed in real and Fourier space, respectively, as (Appendix~\ref{DerivationDifferentialOperator})
\begin{align}
 \mathcal{L}\beta_\mathrm{s}\bm{A}&=\mathcal{M}\left(\mathcal{M}^{-1}\beta_\mathrm{s}\right)\mathcal{M}\bm{A}, \label{LA} \\
 \mathcal{F}\mathcal{L}\beta_\mathrm{s}\bm{A}&=M_K\left(\left(M_K^{-1}\mathcal{F}\beta_\mathrm{s}\right)\ast\left(M_K\mathcal{F}\bm{A}\right)\right), \label{FLA}
\end{align}
where $\mathcal{F}$ denotes the Fourier transformation with respect to $\bm{x}$, ``$\ast$'' the convolution, $\mathcal{M}$ is the volume averaging operator defined by (\ref{M}) or through $\mathcal{M}\bm{A}[\bm{x}]=\frac{1}{V}\int_{|\bm{r}|<R}\bm{A}[\bm{x}+\bm{r}]\d V_r$, and $M_K$ its Fourier transform given by
\begin{equation}
 M_K=\sum_{j=0}^\infty\frac{3(\mathrm{i}K)^{2j}}{(2j+1)!(2j+3)}=\frac{3(\sin K-K\cos K)}{K^3}, \label{Mhat}
\end{equation}
with $K\equiv R|\bm{k}|$ the nondimensionalized absolute value of the wavenumber vector $\bm{k}$ spanning the Fourier space. The effect of $\mathcal{L}$ becomes clear once (\ref{FLA}) is compared with
\begin{equation}
 \mathcal{F}\beta_\mathrm{s}\bm{A}=(\mathcal{F}\beta_\mathrm{s})\ast(\mathcal{F}\bm{A}). \label{FA}
\end{equation}
It can be seen that $\mathcal{L}$ modulates the field $\beta_\mathrm{s}\bm{A}$ in Fourier space via the function $M_K$, which is essentially a low-pass filter that cuts off large-$K$ contributions, while approaching unity in the limit $K\to0$ (figure~\ref{Fig_Mhat}).
Note that $\mathcal{M}^{-1}\beta_\mathrm{s}=nV$, where $n$ is the number density, and, thus, $M_K^{-1}\mathcal{F}\beta_\mathrm{s}=V\mathcal{F}n$, implying $M_K^{-1}\mathcal{F}\beta_\mathrm{s}$ in (\ref{FLA}) is well-defined for values of $\bm{k}$ where $M_K=0$ due to $\mathcal{F}\beta_\mathrm{s}[\bm{k}]=0$.

Choosing $\bm{A}=\rho_\mathrm{f}\D^\mathrm{f}_t\bm{u}_\mathrm{f}-\rho_\mathrm{f}\bm{g}-\beta_\mathrm{f}^{-1}\bm{\nabla}\cdot\bm{\sigma^\mathrm{f}_\mathrm{Re}}+\beta_\mathrm{f}^{-1}\beta_\mathrm{s}\langle\bm{f}_\mathrm{D}\rangle^\mathrm{s}$ in (\ref{FA}) results in the Fourier transform of the closure by \citet{RevilBaudardChauchat13}, whereas the same choice in (\ref{FLA}) yields the Fourier transform of its generalization (\ref{BuoyancyNew2}). Hence, at least for identical spherical particles, the latter is essentially a low-pass-filtered version of the former closure.

\section{Implications for two-fluid models} \label{TwoFluidModels}
\subsection{Linear well-posedness of two-fluid models} \label{Wellposedness}
Linear well-posedness is one of three criteria, next to the existence and uniqueness of a solution, required for a PDE initial-value problem to be well-posed in the sense of Hadamard \citep{PinchoverRubinstein05}. To analyse whether two-fluid models are linearly well-posed when employing our buoyancy closure, we consider a two-fluid model with the closure relations
\begin{align}
 \bm{\nabla}\cdot\bm{\sigma^\mathrm{f}}&=-\bm{\nabla}P_\mathrm{f}+\eta_\mathrm{m}\Delta\bm{u}_\mathrm{m},&\bm{\nabla}\cdot\bm{\sigma^\mathrm{s}}&=-P_\mathrm{s}^\prime\bm{\nabla}\beta_\mathrm{s},&\bm{\sigma^\mathrm{f}_\mathrm{Re}}&=0,&\bm{\sigma^\mathrm{s}_\mathrm{Re}}&=0,&\beta_\mathrm{s}\langle\bm{f}_\mathrm{D}\rangle^\mathrm{s}&=0, \label{Closures}
\end{align}
where $P_\mathrm{f}$ is the effective fluid phase pressure, $P_\mathrm{s}[\beta_\mathrm{s}]$ the $\beta_\mathrm{s}$-dependent effective dispersed-phase pressure \citep[e.g., see][]{Johnsonetal90,Rauter21}, with the physical property $P^\prime_\mathrm{s}\geq0$ (where the prime denotes the derivative), $\eta_\mathrm{m}[\beta_\mathrm{s}]$ the $\beta_\mathrm{s}$-dependent mixture viscosity, and $\bm{u}_\mathrm{m}=\beta_\mathrm{f}\bm{u}_\mathrm{f}+\beta_\mathrm{s}\bm{u}_\mathrm{s}$ the mixture velocity. We choose the form $\bm{\nabla}\cdot\bm{\tau^\mathrm{f}}=\eta_\mathrm{m}\Delta\bm{u}_\mathrm{m}$ as viscous dissipation, since it emerges analytically for particulate Stokes flow \citep{Jackson97} and is also used to model more complex disperse two-phase flows \citep[e.g.,][]{RevilBaudardChauchat13}. However, the more conventional form $\bm{\nabla}\cdot\bm{\tau^\mathrm{f}}=\eta_\mathrm{f}\Delta\bm{u}_\mathrm{f}$ based on only the fluid-phase-averaged velocity $\bm{u}_\mathrm{f}$, with $\eta_\mathrm{f}$ the clear-fluid viscosity, would have the very same consequences and not alter the quality of any of the results presented below. Furthermore, we neglect the Reynolds stresses $\bm{\sigma^\mathrm{f}_\mathrm{Re}}$ and $\bm{\sigma^\mathrm{s}_\mathrm{Re}}$, and the generalized-drag force density $\beta_\mathrm{s}\langle\bm{f}_\mathrm{D}\rangle^\mathrm{s}$ merely to keep the notion simple, since they are usually either quadratic in the flow velocity gradients or come with zero-order derivatives, and therefore do not affect linear well-posedness. An exception are virtual-mass forces, which are addressed in \S\ref{VirtualMass}.

Using (\ref{Closures}) to substitute the respective terms in the buoyancy-improved momentum balances (\ref{MomFluid1}) and (\ref{MomSolid1}) yields
\begin{align}
 \partial_t\beta_\mathrm{f}\rho_\mathrm{f}+\bm{\nabla}\cdot\beta_\mathrm{f}\rho_\mathrm{f}\bm{u}_\mathrm{f}&=0, \label{MassFluidSimple} \\
 \beta_\mathrm{f}\rho_\mathrm{f}\D^\mathrm{f}_t\bm{u}_\mathrm{f}&=-\beta_\mathrm{f}\left(\mathcal{L}\beta_\mathrm{s}\underline{~~}+\mathcal{I}\beta_\mathrm{f}\underline{~~}\right)^{-1}(\bm{\nabla}P_\mathrm{f}-\eta_\mathrm{m}\Delta\bm{u}_\mathrm{m})+\beta_\mathrm{f}\rho_\mathrm{f}\bm{g}, \label{MomFluidSimple} \\
 \partial_t\beta_\mathrm{s}\rho_\mathrm{s}+\bm{\nabla}\cdot\beta_\mathrm{s}\rho_\mathrm{s}\bm{u}_\mathrm{s}&=0, \label{MassSolidSimple} \\
 \beta_\mathrm{s}\rho_\mathrm{s}\D^\mathrm{s}_t\bm{u}_\mathrm{s}+P_\mathrm{s}^\prime\nabla\beta_\mathrm{s}&=-\mathcal{L}\beta_\mathrm{s}\left(\mathcal{L}\beta_\mathrm{s}\underline{~~}+\mathcal{I}\beta_\mathrm{f}\underline{~~}\right)^{-1}(\bm{\nabla}P_\mathrm{f}-\eta_\mathrm{m}\Delta\bm{u}_\mathrm{m})+\beta_\mathrm{s}\rho_\mathrm{s}\bm{g}, \label{MomSolidSimple}
\end{align}
where (\ref{MassFluidSimple}) and (\ref{MassSolidSimple}) are the fluid and dispersed-phase mass balances \citep{Pahtzetal26}. The model is closed by the volume fraction condition $\beta_\mathrm{f}+\beta_\mathrm{s}=1$ and two equations of state, $\rho_\mathrm{s}=\mathrm{const}$ and $\rho_\mathrm{f}=\rho_\mathrm{f}[P_\mathrm{f}]$, with $\rho_\mathrm{f}^\prime\geq0$.

The above model stands as a simple example for the (typically more complex) two-fluid models used in the chemical engineering community \citep[for an overview, see][]{Lhuillieretal13}. It is chosen for demonstration purposes, not for physical realism. In fact, if this model is linearly well-posed, so will be most such models as they usually incorporate more dissipative mechanisms. (Non-dissipative virtual-mass forces are addressed in \S\ref{VirtualMass}.) In the analysis below, we will also take a look at the special cases where $\eta_\mathrm{m}=0$ and/or $P_\mathrm{s}=0$. If both are zero, the model resembles the one-pressure model by \citet{StewartWendroff84}, except for our different buoyancy closure. To transform our model into the latter, one would then only need to apply the small-particle approximation ($\mathcal{L}\mapsto\mathcal{I}$).

To test for linear well-posedness, we consider small-amplitude wave perturbations (subscript ``$\ast$'') of the dependent variables in (\ref{MassFluidSimple})-(\ref{MomSolidSimple}) around a base state (subscript ``$0$''):
\begin{align}
 &\beta_\mathrm{s}=\beta_{\mathrm{s}0}+\beta_{\mathrm{s}\ast},&&\bm{u}_\mathrm{s}=\bm{u}_{\mathrm{s}0}+\bm{u}_{\mathrm{s}\ast},&&\bm{u}_\mathrm{f}=\bm{u}_{\mathrm{f}0}+\bm{u}_{\mathrm{f}\ast},&&P_\mathrm{f}=P_{\mathrm{f}0}+P_{\mathrm{f}\ast}.
\end{align}
In the limit of infinitely large wave numbers and frequencies of these perturbation, relevant for a Hadamard instability, the base state, which varies over finite characteristic length and time, can be assumed to be locally constant \citep{JosephSaut90}. Then, linearizing (\ref{MassFluidSimple})-(\ref{MomSolidSimple}) in these perturbations yields
\begin{align}
 -\rho_{\mathrm{f}0}\partial_t\beta_{\mathrm{s}\ast}-\rho_{\mathrm{f}0}\bm{u}_{\mathrm{f}0}\cdot\bm{\nabla}\beta_{\mathrm{s}\ast}+\beta_{\mathrm{f}0}\rho_{\mathrm{f}0}\bm{\nabla}\cdot\bm{u}_{\mathrm{f}\ast}&=-\beta_{\mathrm{f}0}\rho_{\mathrm{f}0}^\prime\partial_tP_{\mathrm{f}\ast}-\beta_{\mathrm{f}0}\rho_{\mathrm{f}0}^\prime\bm{u}_{\mathrm{f}0}\cdot\bm{\nabla}P_{\mathrm{f}\ast}, \label{MassFluidSimplelin} \\
 \beta_{\mathrm{f}0}\rho_{\mathrm{f}0}\partial_t\bm{u}_{\mathrm{f}\ast}+\beta_{\mathrm{f}0}\rho_{\mathrm{f}0}\bm{u}_{\mathrm{f}0}\cdot\bm{\nabla}\bm{u}_{\mathrm{f}\ast}&=-\beta_{\mathrm{f}0}\mathcal{N}(\bm{\nabla}P_{\mathrm{f}\ast}-\eta_{\mathrm{m}0}\Delta\bm{u}_{\mathrm{m}1\ast})+(\beta_\mathrm{f}\rho_\mathrm{f})_1\bm{g}, \label{MomFluidSimplelin} \\
 \rho_\mathrm{s}\partial_t\beta_{\mathrm{s}\ast}+\rho_\mathrm{s}\bm{u}_{\mathrm{s}0}\cdot\bm{\nabla}\beta_{\mathrm{s}\ast}+\beta_{\mathrm{s}0}\rho_\mathrm{s}\bm{\nabla}\cdot\bm{u}_{\mathrm{s}\ast}&=0, \label{MassSolidSimplelin} \\
 \beta_{\mathrm{s}0}\rho_\mathrm{s}\partial_t\bm{u}_{\mathrm{s}\ast}+\beta_{\mathrm{s}0}\rho_\mathrm{s}\bm{u}_{\mathrm{s}0}\cdot\bm{\nabla}\bm{u}_{\mathrm{s}\ast}+P_{\mathrm{s}0}^\prime\nabla\beta_{\mathrm{s}\ast}&=-\beta_{\mathrm{s}0}\mathcal{M}^2\mathcal{N}(\bm{\nabla}P_{\mathrm{f}\ast}-\eta_{\mathrm{m}0}\Delta\bm{u}_{\mathrm{m}1\ast})+\beta_\mathrm{s}\rho_\mathrm{s}\bm{g}, \label{MomSolidSimplelin}
\end{align}
where $\rho_{\mathrm{f}0}\equiv\rho_\mathrm{f}[P_{\mathrm{f}0}]$, $\rho_{\mathrm{f}0}^\prime\equiv\rho_\mathrm{f}^\prime[P_{\mathrm{f}0}]$, $\beta_{\mathrm{f}0}\equiv1-\beta_{\mathrm{s}0}$, $P_{\mathrm{s}0}^\prime\equiv P_\mathrm{s}^\prime[\beta_{\mathrm{s}0}]$, $\eta_{\mathrm{m}0}\equiv\eta_\mathrm{m}[\beta_{\mathrm{s}0}]$, $\mathcal{N}\equiv(\beta_{\mathrm{s}0}\mathcal{M}^2+\beta_{\mathrm{f}0}\mathcal{I})^{-1}$, $\bm{u}_{\mathrm{m}1\ast}\equiv\beta_{\mathrm{f}0}\bm{u}_{\mathrm{f}\ast}+\beta_{\mathrm{s}0}\bm{u}_{\mathrm{s}\ast}-(\bm{u}_{\mathrm{f}0}-\bm{u}_{\mathrm{s}0})\beta_{\mathrm{s}\ast}$, and $(\beta_\mathrm{f}\rho_\mathrm{f})_1$ denotes the linearization of $\beta_\mathrm{f}\rho_\mathrm{f}$. Note that we used (\ref{LA}), assuming identically-sized spheres for simplicity (discussed at the end of this section). Now, we Fourier-transform (\ref{MassFluidSimplelin})-(\ref{MomSolidSimplelin}) in space and time---formally, $\bm{\nabla}\mapsto\mathrm{i}\bm{k}$, $\mathcal{M}\mapsto M_K$, $\mathcal{N}\mapsto N_K\equiv1/(\beta_{\mathrm{s}0}M_K^2+\beta_{\mathrm{f}0})$, and $\partial_t\mapsto-\mathrm{i}\Omega$, where $\Omega$ is the wave frequency---and keep only the terms proportional to $\Omega$, $k\equiv|\bm{k}|$, and $k^2$; terms of $O(1)$, which can potentially also cause Hadamard instabilities, will be considered indirectly via analysis of the dimensionality of the solution space (explained shortly). This formally implies $M_K\mapsto0$, $N_K\mapsto1/\beta_{\mathrm{f}0}$, and $\bm{g}\mapsto0$ in the Fourier transforms of (\ref{MomFluidSimplelin}) and (\ref{MomSolidSimplelin}). Put together, these steps yield
\begin{align}
 \beta_{\mathrm{f}0}\rho_{\mathrm{f}0}^\prime(\Omega-\bm{u}_{\mathrm{f}0}\cdot\bm{k})\hat P_{\mathrm{f}\ast}-\beta_{\mathrm{f}0}\rho_{\mathrm{f}0}\bm{k}\cdot\bm{\hat u}_{\mathrm{f}\ast}-\rho_{\mathrm{f}0}(\Omega-\bm{u}_{\mathrm{f}0}\cdot\bm{k})\hat\beta_{\mathrm{s}\ast}&=0, \label{MassFluidSimplelinFourier} \\
 -\hat P_{\mathrm{f}\ast}\bm{k}+\beta_{\mathrm{f}0}\rho_{\mathrm{f}0}(\Omega-\bm{u}_{\mathrm{f}0}\cdot\bm{k})\bm{\hat u}_{\mathrm{f}\ast}+\mathrm{i}\eta_{\mathrm{m}0}k^2\bm{\hat u}_{\mathrm{m}1\ast}&=0, \label{MomFluidSimplelinFourier} \\
 \rho_\mathrm{s}(\Omega-\bm{u}_{\mathrm{s}0}\cdot\bm{k})\hat\beta_{\mathrm{s}\ast}-\beta_{\mathrm{s}0}\rho_\mathrm{s}\bm{k}\cdot\bm{\hat u}_{\mathrm{s}\ast}&=0, \label{MassSolidSimplelinFourier} \\
 -P_{\mathrm{s}0}^\prime\hat\beta_{\mathrm{s}\ast}\bm{k}+\beta_{\mathrm{s}0}\rho_\mathrm{s}(\Omega-\bm{u}_{\mathrm{s}0}\cdot\bm{k})\bm{\hat u}_{\mathrm{s}\ast}&=0, \label{MomSolidSimplelinFourier}
\end{align}
where the hat indicates the Fourier transformation with respect to $\bm{x}$ and $t$. To simplify the problem, we decompose $\bm{\hat u}_{\mathrm{f}\ast}$ and $\bm{\hat u}_{\mathrm{s}\ast}$ into longitudinal perturbations parallel to $\bm{k}$ (superscript ``$\parallel$'') and transverse perturbations normal to $\bm{k}$ (superscript ``$\perp$''),
\begin{align}
 \bm{\hat u}^\parallel_{\mathrm{f}\ast}&\equiv\bm{\hat u}_{\mathrm{f}\ast}\cdot\bm{e_k}\bm{e_k}\equiv\hat u^\parallel_{\mathrm{f}\ast}\bm{e_k},&\bm{\hat u}^\parallel_{\mathrm{s}\ast}&\equiv\bm{\hat u}_{\mathrm{s}\ast}\cdot\bm{e_k}\bm{e_k}\equiv\hat u^\parallel_{\mathrm{s}\ast}\bm{e_k}, \\
 \bm{\hat u}^\perp_{\mathrm{f}\ast}&\equiv\bm{\hat u}_{\mathrm{f}\ast}-\bm{\hat u}^\parallel_{\mathrm{f}\ast},&\bm{\hat u}^\perp_{\mathrm{s}\ast}&\equiv\bm{\hat u}_{\mathrm{s}\ast}-\bm{\hat u}^\parallel_{\mathrm{s}\ast},
\end{align}
where $\bm{e_k}=\bm{k}/k$ is the unit vector in the direction of $\bm{k}$. Furthermore, without loss of generality, we use a Cartesian coordinate system with one of its axes aligned with $\bm{e_k}$. Defining
\begin{equation}
\bm{X}\equiv\left(\hat P_{\mathrm{f}\ast},\hat u^\parallel_{\mathrm{f}\ast},\hat u^\perp_{\mathrm{f}1\ast},\hat u^\perp_{\mathrm{f}2\ast},\hat\beta_{\mathrm{s}\ast},\hat u^\parallel_{\mathrm{s}\ast},\hat u^\perp_{\mathrm{s}1\ast},\hat u^\perp_{\mathrm{s}2\ast}\right)^\mathrm{T},
\end{equation}
where the indices $1$ and $2$ indicate the two orthogonal transverse directions, we can therefore write (\ref{MassFluidSimplelinFourier})-(\ref{MomSolidSimplelinFourier}) as
\begin{equation}
 \bm{C}\bm{X}=0. \label{PDELinear}
\end{equation}
Here $\bm{C}$ is a $8\times8$ matrix given by
\begin{equation}
 \bm{C}\equiv
 \begin{pNiceMatrix}[margin]
  \bm{C}^\parallel_\mathrm{f} & \bm{0} & \Block{2-2}<\Large>{\bm{C}^\mathrm{s}_\mathrm{f}} & \\
	\bm{0} & \bm{C}^\perp_\mathrm{f} & & \\
	\bm{0} & \bm{0} & \bm{C}^\parallel_\mathrm{s} & \bm{0} \\
	\bm{0} & \bm{0} & \bm{0} & \bm{C}^\perp_\mathrm{s}
 \end{pNiceMatrix},
\end{equation}
with $\bm{C}^\parallel_\mathrm{f}$, $\bm{C}^\perp_\mathrm{f}$, $\bm{C}^\parallel_\mathrm{s}$, and $\bm{C}^\perp_\mathrm{s}$ $2\times2$ matrices defined, respectively, as
\begin{align}
 \bm{C}^\parallel_\mathrm{f}&\equiv
 \begin{pmatrix}
  \beta_{\mathrm{f}0}\rho_{\mathrm{f}0}^\prime\left(\Omega-u^\parallel_{\mathrm{f}0}k\right) & -\beta_{\mathrm{f}0}\rho_{\mathrm{f}0}k \\
  -k & \beta_{\mathrm{f}0}\rho_{\mathrm{f}0}\left(\Omega-u^\parallel_{\mathrm{f}0}k+\mathrm{i}\eta_\mathrm{m}k^2/\rho_{\mathrm{f}0}\right)
 \end{pmatrix}, \\
 \bm{C}^\perp_\mathrm{f}&\equiv
 \begin{pmatrix}
  \beta_{\mathrm{f}0}\rho_{\mathrm{f}0}\left(\Omega-u^\parallel_{\mathrm{f}0}k+\mathrm{i}\eta_\mathrm{m}k^2/\rho_{\mathrm{f}0}\right) & 0 \\
  0 & \beta_{\mathrm{f}0}\rho_{\mathrm{f}0}\left(\Omega-u^\parallel_{\mathrm{f}0}k+\mathrm{i}\eta_\mathrm{m}k^2/\rho_{\mathrm{f}0}\right)
 \end{pmatrix}, \\
 \bm{C}^\parallel_\mathrm{s}&\equiv
 \begin{pmatrix}
  \rho_\mathrm{s}\left(\Omega-u^\parallel_{\mathrm{s}0}k\right) & -\beta_{\mathrm{s}0}\rho_\mathrm{s}k \\
  -P_{\mathrm{s}0}^\prime k & \beta_{\mathrm{s}0}\rho_\mathrm{s}\left(\Omega-u^\parallel_{\mathrm{s}0}k\right)
 \end{pmatrix}, \\
 \bm{C}^\perp_\mathrm{s}&\equiv
 \begin{pmatrix}
  \beta_{\mathrm{s}0}\rho_\mathrm{s}\left(\Omega-u^\parallel_{\mathrm{s}0}k\right) & 0 \\
  0 & \beta_{\mathrm{s}0}\rho_\mathrm{s}\left(\Omega-u^\parallel_{\mathrm{s}0}k\right)
 \end{pmatrix},
\end{align}
and $\bm{C}^\mathrm{s}_\mathrm{f}$ a $4\times4$ matrix defined as
\begin{equation}
 \bm{C}^\mathrm{s}_\mathrm{f}\equiv
 \begin{pmatrix}
  -\rho_{\mathrm{f}0}\left(\Omega-u^\parallel_{\mathrm{f}0}k\right) & 0 & 0 & 0 \\
  -\mathrm{i}\eta_{\mathrm{m}0}k^2\left(u^\parallel_{\mathrm{f}0}-u^\parallel_{\mathrm{s}0}\right) & \mathrm{i}\beta_{\mathrm{s}0}\eta_{\mathrm{m}0}k^2 & 0 & 0 \\
  -\mathrm{i}\eta_{\mathrm{m}0}k^2\left(u^\perp_{\mathrm{f}10}-u^\perp_{\mathrm{s}10}\right) & 0 & \mathrm{i}\beta_{\mathrm{s}0}\eta_{\mathrm{m}0}k^2 & 0 \\
  -\mathrm{i}\eta_{\mathrm{m}0}k^2\left(u^\perp_{\mathrm{f}20}-u^\perp_{\mathrm{s}20}\right) & 0 & 0 & \mathrm{i}\beta_{\mathrm{s}0}\eta_{\mathrm{m}0}k^2
 \end{pmatrix},
\end{equation}
where the subscript ``$0$'' indicates again the base state value of a quantity; for example, $u^\perp_{\mathrm{f}10}$ is the base state value of $u^\perp_{\mathrm{f}1}$, the first transverse component of $\bm{u}_\mathrm{f}$.

We now look for those frequencies $\Omega$ that allow for non-trivial perturbation vectors $\bm{X}\ne0$ in (\ref{PDELinear}), which can be obtained from the condition
\begin{equation}
 \det(\bm{C})=\det(\bm{C}^\parallel_\mathrm{f})\det(\bm{C}^\perp_\mathrm{f})\det(\bm{C}^\parallel_\mathrm{s})\det(\bm{C}^\perp_\mathrm{s})=0. \label{detC=0}
\end{equation}
It is equivalent to the condition
\begin{equation}
 \det(\bm{C}^\parallel_\mathrm{f})=0\lor\det(\bm{C}^\perp_\mathrm{f})=0\lor\det(\bm{C}^\parallel_\mathrm{s})=0\lor\det(\bm{C}^\perp_\mathrm{s})=0 \label{detC=0_2}
\end{equation}
and yields eight corresponding frequency solutions:
\begin{align}
 \eta_{\mathrm{m}0}&=0:&\Omega^\parallel_{\mathrm{f}1[2]}&=\left(u^\parallel_{\mathrm{f}0}+[-]\left(\beta_{\mathrm{f}0}\rho_{\mathrm{f}0}^\prime\right)^{-1/2}\right)k,&\Omega^\perp_{\mathrm{f}1,2}&=u^\parallel_{\mathrm{f}0}k, \label{OmegaSolutions1} \\
 \eta_{\mathrm{m}0}&>0:&\Omega^\parallel_{\mathrm{f}1}&\simeq u^\parallel_{\mathrm{f}0}k,&\Omega^\parallel_{\mathrm{f}2},\Omega^\perp_{\mathrm{f}1,2}&\simeq-\mathrm{i}\eta_{\mathrm{m}0}k^2/\rho_{\mathrm{f}0}, \\
 \eta_{\mathrm{m}0}&\geq0:&\Omega^\parallel_{\mathrm{s}1[2]}&=\left(u^\parallel_{\mathrm{s}0}+[-]\left(P_{\mathrm{s}0}^\prime/\rho_\mathrm{s}\right)^{1/2}\right)k,&\Omega^\perp_{\mathrm{s}1,2}&=u^\parallel_{\mathrm{s}0}k. \label{OmegaSolutions3}
\end{align}
Here the symbol ``$\simeq$'' denotes the asymptotic expansion in the limit $k\to\infty$, the subscript ``$1,2$'' indicates a double solution, while the subscript ``$1[2]$'' means that the second solution refers to the symbol surrounded by square brackets on the corresponding right-hand side, whereas the first solution refers to the symbol immediately before the square brackets. Note that, for an incompressible fluid, $\rho^\prime_{\mathrm{f}0}=0$, the determinant $\det(\bm{C}^\parallel_\mathrm{f})=-\beta_{\mathrm{f}0}\rho_{\mathrm{f}0}k^2$ is independent of $\Omega$ and strictly smaller than zero, implying that the double solution $\Omega^\parallel_{\mathrm{f}1,2}$ no longer exist, while the other solutions remain unaffected.

Due to $\rho_{\mathrm{f}0}^\prime\geq0$ and $P^\prime_\mathrm{s}\geq0$, all the above frequency solutions are asymptotically either negative imaginary, associated with damping, or real-valued, associated with pure oscillations, implying that the $O(k)$ and $O(k^2)$ terms themselves are unable to cause a Hadamard instability. However, the system of PDEs is not necessarily linearly well-posed, since the previously neglected $O(1)$ terms can conspire with the $O(k)$ terms to cause unbounded growth as $k^b$, with $0<b<1$ \citep{Langhametal25b}. This happens when $n\geq2$ real-valued frequency solutions asymptotically coincide for a given base state and their corresponding nullspaces (i.e., the spaces of the non-trivial perturbation vector solutions) are less than $n$-dimensional \citep{Langhametal25b}. To check for this possibility we calculate the basis vectors of the nullspaces corresponding to the frequency solutions in (\ref{OmegaSolutions1})-(\ref{OmegaSolutions3}):
\begin{align}
 \eta_{\mathrm{m}0}&=0:&\bm{X}^\parallel_{\mathrm{f}1[2]}&=\left(+[-]\rho_{\mathrm{f}0}\left(\beta_{\mathrm{f}0}/\rho_{\mathrm{f}0}^\prime\right)^{1/2},1,0,0,0,0,0,0\right)^\mathrm{T}, \\
 \eta_{\mathrm{m}0}&>0:&\bm{X}^\parallel_{\mathrm{f}1[2]}&\simeq\left(1[0],0[1],0,0,0,0,0,0\right)^\mathrm{T}, \\
 \eta_{\mathrm{m}0}&\geq0:&\bm{X}^\perp_{\mathrm{f}1[2]}&=\left(0,0,1[0],0[1],0,0,0,0\right)^\mathrm{T}, \\
 \eta_{\mathrm{m}0}&=0:&\bm{X}^\parallel_{\mathrm{s}1[2]}&=\left(\beta_{\mathrm{f}0}\rho_{\mathrm{f}0}\left(u^\parallel_{\mathrm{s}0}+[-]\left(P_{\mathrm{s}0}^\prime/\rho_\mathrm{s}\right)^{1/2}-u^\parallel_{\mathrm{f}0}\right),1,0,\right. \nonumber \\
 &&&\quad\;\;\left.0,\chi_{1[2]},+[-]\beta_{\mathrm{s}0}^{-1}\left(P_{\mathrm{s}0}^\prime/\rho_\mathrm{s}\right)^{1/2}\chi_{1[2]},0,0\right)^\mathrm{T}, \\
 \eta_{\mathrm{m}0}&>0:&\bm{X}^\parallel_{\mathrm{s}1[2]}&\simeq\left(0,\beta_{\mathrm{s}0}\left(u^\parallel_{\mathrm{f}0}-u^\parallel_{\mathrm{s}0}-[+]\left(P_{\mathrm{s}0}^\prime/\rho_\mathrm{s}\right)^{1/2}\right),\beta_{\mathrm{s}0}\left(u^\perp_{\mathrm{f}10}-u^\perp_{\mathrm{s}10}\right),\right. \nonumber \\
 &&&\quad\;\;\left.\beta_{\mathrm{s}0}\left(u^\perp_{\mathrm{f}20}-u^\perp_{\mathrm{s}20}\right),\beta_{\mathrm{f}0}\beta_{\mathrm{s}0},+[-]\beta_{\mathrm{f}0}\left(P_{\mathrm{s}0}^\prime/\rho_\mathrm{s}\right)^{1/2},0,0\right)^\mathrm{T}, \\
 \eta_{\mathrm{m}0}&=0:&\bm{X}^\perp_{\mathrm{s}1[2]}&=(0,0,0,0,0,0,1[0],0[1])^\mathrm{T}, \\
 \eta_{\mathrm{m}0}&>0:&\bm{X}^\perp_{\mathrm{s}1[2]}&\simeq(0,0,-\beta_{\mathrm{s}0}/\beta_{\mathrm{f}0}[0],0[-\beta_{\mathrm{s}0}/\beta_{\mathrm{f}0}],0,0,1[0],0[1])^\mathrm{T},
\end{align}
where
\begin{equation}
 \chi_{1[2]}\equiv-\beta_{\mathrm{f}0}\left(\left(u^\parallel_{\mathrm{s}0}+[-]\left(P_{\mathrm{s}0}^\prime/\rho_\mathrm{s}\right)^{1/2}-u^\parallel_{\mathrm{f}0}\right)^{-1}-\beta_{\mathrm{f}0}\rho_{\mathrm{f}0}^\prime\left(u^\parallel_{\mathrm{s}0}+[-]\left(P_{\mathrm{s}0}^\prime/\rho_\mathrm{s}\right)^{1/2}-u^\parallel_{\mathrm{f}0}\right)\right).
\end{equation}
One immediately obtains that the basis vectors corresponding to the transverse modes, $\bm{X}^\perp_{\mathrm{f}1}$, $\bm{X}^\perp_{\mathrm{f}2}$, $\bm{X}^\perp_{\mathrm{s}1}$, and $\bm{X}^\perp_{\mathrm{s}2}$, are always linearly independent from each other and from the basis vectors corresponding to the longitudinal modes, i.e., they never cause problems. Furthermore, we can infer the following statements:
\begin{itemize}
 \item If variations of the effective dispersed-phase pressure are neglected, $P^\prime_\mathrm{s}=0$, then $\Omega^\parallel_{\mathrm{s}1}=\Omega^\parallel_{\mathrm{s}2}$ and $\bm{X}^\parallel_{\mathrm{s}1}=\bm{X}^\parallel_{\mathrm{s}2}$ regardless of the behaviour of $\eta_\mathrm{m}$. Hence, the nullspace corresponding to $\Omega^\parallel_{\mathrm{s}1}$ and $\Omega^\parallel_{\mathrm{s}2}$ is just one-dimensional rather than two-dimensional, implying that the system of PDEs is ill-posed due to unbounded growth as $k^{1/2}$ \citep{Langhametal25b}.
 \item If $P_\mathrm{s}^\prime>0$ and if the fluid is incompressible, $\rho_{\mathrm{f}0}^\prime=0$, then the system is linearly well-posed regardless of the behaviour of $\eta_\mathrm{m}$, since the frequency solutions $\Omega^\parallel_{\mathrm{f}1}$ and $\Omega^\parallel_{\mathrm{f}2}$ and their corresponding basis vectors $\bm{X}^\parallel_{\mathrm{f}1}$ and $\bm{X}^\parallel_{\mathrm{f}2}$ no longer exist, while the remaining six basis vectors are always linearly independent from each other.
 \item If $P_\mathrm{s}^\prime>0$ and if the fluid is compressible, $\rho_{\mathrm{f}0}^\prime>0$, then the system is linearly well-posed for $\eta_\mathrm{m}>0$ but generally ill-posed if viscosity effects are neglected, $\eta_\mathrm{m}=0$. In the latter case, for those base states that satisfy $\Omega^\parallel_{\mathrm{f}1}=\Omega^\parallel_{\mathrm{s}1}$ or $\Omega^\parallel_{\mathrm{f}1}=\Omega^\parallel_{\mathrm{s}2}$ or $\Omega^\parallel_{\mathrm{f}2}=\Omega^\parallel_{\mathrm{s}1}$ or $\Omega^\parallel_{\mathrm{f}2}=\Omega^\parallel_{\mathrm{s}2}$, the corresponding basis vectors coincide, i.e., $\bm{X}^\parallel_{\mathrm{f}1}=\bm{X}^\parallel_{\mathrm{s}1}$ or $\bm{X}^\parallel_{\mathrm{f}1}=\bm{X}^\parallel_{\mathrm{s}2}$ or $\bm{X}^\parallel_{\mathrm{f}2}=\bm{X}^\parallel_{\mathrm{s}1}$ or $\bm{X}^\parallel_{\mathrm{f}2}=\bm{X}^\parallel_{\mathrm{s}2}$, respectively, implying again unbounded growth as $k^{1/2}$ \citep{Langhametal25b}. By contrast, viscous dissipation for $\eta_\mathrm{m}>0$ prevents such resonances from occurring through asymptotically decoupling fluid phase and dispersed-phase oscillations from one another. 
\end{itemize}
Note that, if we modelled viscous dissipation as $\bm{\nabla}\cdot\bm{\tau^\mathrm{f}}=\eta_\mathrm{f}\Delta\bm{u}_\mathrm{f}$ instead of $\bm{\nabla}\cdot\bm{\tau^\mathrm{f}}=\eta_\mathrm{m}\Delta\bm{u}_\mathrm{m}$, the above statements would remain true if $\eta_\mathrm{m}$ were formally replaced by $\eta_\mathrm{f}$.

In summary, even simplistic two-fluid models are linearly well-posed when employing our buoyancy closure. One only requires standard viscous damping of the fluid phase and an effective dispersed-phase pressure $P_\mathrm{s}$ that strictly monotonously increases with the dispersed-phase volume fraction $\beta_\mathrm{s}$, which is a requirement dictated by physics \citep[e.g., granular kinetic theory,][]{Berzi24}.

In the above analysis, we have argued on the basis of (\ref{LA}), which is valid only for systems of identically-sized spheres. However, it is straightforward to generalize this analysis to arbitrary particle sizes and shapes, giving the same results. This is because the very definition of the operator $\mathcal{L}$ in (\ref{L}) always means that spatial variations smaller than the characteristic particle size are smoothed out, attenuating buoyancy effects as $k^{-4}$ when $k\to\infty$.

\subsubsection{Virtual-mass force density} \label{VirtualMass}
In addition to buoyancy, virtual-mass-related effects have also been linked to the ill-posedness of two-fluid models \citep{Lhuillieretal13}. A standard expression for the virtual-mass force density $\beta_\mathrm{s}\langle\bm{f}_\mathrm{v}\rangle^\mathrm{s}\equiv\beta_\mathrm{s}\langle\sum_pX^p\bm{F}^p_\mathrm{v}/V^p\rangle^\mathrm{s}$, where $\bm{F}^p_\mathrm{v}$ is the virtual-mass force acting on a particle $p$, is that proposed by \citet{Jackson00}:
\begin{equation}
 \beta_\mathrm{s}\left\langle\bm{f}_\mathrm{v}\right\rangle^\mathrm{s}=C_\mathrm{v}\rho_\mathrm{f}\beta_\mathrm{s}\beta_\mathrm{f}\left(\D^\mathrm{f}_t\bm{u}_\mathrm{f}-\D^\mathrm{s}_t\bm{u}_\mathrm{s}\right), \label{fVMJackson}
\end{equation}
where $C_\mathrm{v}[\beta_\mathrm{s}]$ is the virtual-mass coefficient, with $C_\mathrm{v}[0]=1/2$. We have written (\ref{fVMJackson}) in a form that involves multiplication with $\beta_\mathrm{f}$, analogous to the generalized-drag force density expression in \S\ref{StressFluidParticleInteractions}, in order to highlight that $\beta_\mathrm{s}\left\langle\bm{f}_\mathrm{v}\right\rangle^\mathrm{s}$ is a force per unit volume of the fluid-particle mixture. Incorporating (\ref{fVMJackson}) as an interaction force density in (\ref{MomFluidSimple}) and (\ref{MomSolidSimple}) results in frequency solutions $\Omega\sim k$ with positive imaginary part for many base states and therefore in growth as $k^1$ and ill-posedness. However, if one took into account that $\bm{u}_\mathrm{f}$, which plays a role analogous to the background flow velocity $\bm{\tilde u}$ in the MRG equation (\ref{MRG}), should actually be averaged over the particles' volumes (e.g., see (\ref{MRG})), one should model $\beta_\mathrm{s}\left\langle\bm{f}_\mathrm{v}\right\rangle^\mathrm{s}$ as 
\begin{equation}
 \beta_\mathrm{s}\left\langle\bm{f}_\mathrm{v}\right\rangle^\mathrm{s}=C_\mathrm{v}\rho_\mathrm{f}\beta_\mathrm{f}\left(\mathcal{L}\beta_\mathrm{s}\D^\mathrm{f}_t\bm{u}_\mathrm{f}-\beta_\mathrm{s}\D^\mathrm{s}_t\bm{u}_\mathrm{s}\right). \label{fVM}
\end{equation}
This becomes immediately clear when comparing with the manner in which the average generalized-buoyancy force on a particle $p$, $\langle\bm{F}^p_\mathrm{B}\rangle=\int_{\mathbb{V}^p}\bm{\nabla}\cdot\bm{\hat\sigma^\mathrm{f}}\d V$ is linked to the corresponding generalized-buoyancy force density $\beta_\mathrm{s}\left\langle\bm{f}_\mathrm{B}\right\rangle^\mathrm{s}$ in (\ref{BuoyancyNew0}). Incorporating (\ref{fVM}) as an interaction force density in (\ref{MomFluidSimple}) and (\ref{MomSolidSimple}) does not alter linear well-posedness, since all above frequency solutions and basis vectors remain the same if the dispersed-phase mass density $\rho_\mathrm{s}$ is formally replaced as
\begin{equation}
 \rho_\mathrm{s}\mapsto\rho_\mathrm{s}+C_{\mathrm{v}0}\rho_{\mathrm{f}0}\beta_{\mathrm{f}0},
\end{equation}
where $C_{\mathrm{v}0}\equiv C_\mathrm{v}[\beta_{\mathrm{s}0}]$.

\subsection{Numerical implementation in two-fluid models} \label{TwoFluidNumericalImplementation}
The relevance of our improved buoyancy closure for numerical two-fluid models depends on the size of the mesh grid cells discretizing a given system. Since a value of a physical quantity assigned to a grid cell already represents an average over the cell domain, the effect of the smoothing operator $\mathcal{L}$ distinguishing our closure from its small-particle approximation, given by the closure by \citet{RevilBaudardChauchat13}, (\ref{BuoyancyChauchat}), is negligible if the grid size is much larger than $R$. Hence, for simulations of large-scale systems with two fluid models, it suffices to employ this small-particle approximation. However, if the grid size is of the order of $R$ or smaller, the effects of $\mathcal{L}$ become relevant. Some researchers are under the impression that it makes no physical sense to analyse two-fluid continuum models at such small scales. For example, \citet[][pp.~112-115]{Jackson00}, when analysing the stability of homogeneous fluidized beds, stated that, for wave disturbances of the macroscopic flow quantities with wavelengths of the order of $R$, ``continuum equations are clearly inappropriate.'' However, we disagree with this perception. When deriving two-fluid continuum equations via ensemble averaging the microscopic equations of motion of the fluid and particles, no separation between the sizes of the microscopic and macroscopic scales is mathematically required \citep{Prosperetti98,Pahtzetal26}. More importantly, there are, indeed, disperse two-phase flows where macroscopic quantities change over distances (much) smaller than $R$. A notable example is fluid-driven sediment transport, which is characterized by a typically sudden transition from a densely packed bed to a potentially very dilute transport layer~\citep{Chassagneetal23,Tholenetal23,Fryetal24}, e.g., see figures~\ref{ConcentrationVelocityProfiles} and \ref{Fig_SigmaProfileszx}.

When numerical mesh grid cells are of the order of $R$ or smaller, we suggest to use the form of the two-fluid momentum balances in (\ref{MomFluid2}) and (\ref{MomSolid2}) to implement our buoyancy closure in two-fluid models. This form involves no modification of the effective fluid phase stress tensor $\bm{\sigma^\mathrm{f}}$, while the operator $\mathcal{L}$ is applied on only the single term $\beta_\mathrm{s}\bm{A}$, with $\bm{A}\equiv\rho_\mathrm{f}\D^\mathrm{f}_t\bm{u}_\mathrm{f}-\rho_\mathrm{f}\bm{g}-\beta_\mathrm{f}^{-1}\bm{\nabla}\cdot\bm{\sigma^\mathrm{f}_\mathrm{Re}}+\beta_\mathrm{f}^{-1}\beta_\mathrm{s}\langle\bm{f}_\mathrm{D}\rangle^\mathrm{s}$. Then, despite the mathematical complexity involved in $\mathcal{L}$, there is a simple approximate manner with which its effects can be accounted for:
\begin{equation}
 \mathcal{L}\beta_\mathrm{s}\bm{A}\approx\beta_\mathrm{s}\mathcal{M}^2\bm{A}, \label{LApproximation}
\end{equation}
where the radius $R$ involved in the definition of $\mathcal{M}$ should, in the case of arbitrarily sized and shaped particles, be interpreted as the volume-equivalent radius corresponding to the average particle volume. The approximation (\ref{LApproximation}) preserves the regularization effect of $\mathcal{L}$ and, in the case of identically-sized spheres, agrees with the exact expression for both extreme limits in which either of the two quantities $\beta_\mathrm{s}$ and $\bm{A}$ spatially varies much more than the other one. The numerical implementation of $\mathcal{M}^2\bm{A}$ is simple. Presume $\bm{A}$ is given for every numerical grid cell $\bm{x}_i$. Then, assign to each grid cell the value of $\bm{A}$ averaged over a ball of radius $R$ centred at $\bm{x}_i$, giving $\mathcal{M}\bm{A}[\bm{x}_i]$. Repeating the procedure for $\mathcal{M}\bm{A}[\bm{x}_i]$ results in $\mathcal{M}^2\bm{A}[\bm{x}_i]$.

\section{Summary and Conclusions}
In this study, we have revisited an old problem that was the subject of debate several decades ago, and which has never been fully resolved: What is the mathematical closure that governs the generalized-buoyancy force density in general disperse two-phase flow? To answer this question, we have used rigorous mathematical analysis, physical intuition, particle-resolving numerical simulations, and thought experiments. They have demonstrated that buoyancy is caused by the effective mechanical fluid phase stress $\bm{\sigma^\mathrm{f}}\equiv\beta_\mathrm{f}\langle\bm{\sigma}\rangle^\mathrm{f}+\bm{\sigma^\mathrm{f}_\mathrm{s}}$, which includes as a contribution the pseudo-stress $\bm{\sigma^\mathrm{s}}$ that emerges due to fluid-particle interactions (\S\ref{PseudoStresses}). This result supports the buoyancy closure by \citet{RevilBaudardChauchat13}, (\ref{BuoyancyChauchat}), but strongly challenges the closures by \citet{Zhangetal07b}, (\ref{BuoyancyZhang}), and the ``Jackson closure'', (\ref{BuoyancyJackson}), by \citet{AndersonJackson67} and \citet{Jackson00}. In particular, (\ref{BuoyancyJackson}), which is often used in the chemical engineering community \citep[e.g.,][]{Jamshidietal21}, is clearly falsified by particle-resolving simulations of a single small spherical particle in a pressure-driven turbulent free-surface flow (\S\ref{HorizontalSettling}). These simulations show that, beyond any reasonable doubt, gradients of the background fluid phase Reynolds shear stress, unlike predicted by (\ref{BuoyancyJackson}), do not contribute to the generalized-buoyancy force on the particle.

Due to the above, our results support one side of an ongoing dispute about the direction of the generalized-buoyancy force in turbulent sediment transport \citep[e.g.,][]{ChiewParker94,Christensen95,Chenetal10,Maurinetal18}. The found absence of Reynolds stress gradient contributions implies that, for steady, uniform conditions at low dispersed-phase volume fraction $\beta_\mathrm{s}$, the average generalized-buoyancy force $\langle\bm{F}_\mathrm{B}\rangle$ has only a component in the direction normal to the flow \citep{Maurinetal18}. Hence, our results strongly challenge the physics behind standard models on the bed slope dependence of the sediment transport threshold \citep[e.g.,][]{ChiewParker94,Chenetal10}, which were made under the assumption that buoyancy and gravity are aligned. They also suggest that the slope dependencies of the drag coefficient observed by \citet{Lambetal17} for partially submerged objects in a slope-driven stream when neglecting Reynolds stress contributions to the buoyancy force (M.~P.~Lamb, personal communication 2024), as discussed in \S\ref{Introduction}, are a real occurrence rather than evidence for a mismodeling of buoyancy.

The closure by \citet{RevilBaudardChauchat13}, (\ref{BuoyancyChauchat}), is valid only in the small-particle approximation. However, this approximation causes an inconsistency for some fluid-particle systems at complete rest (\S\ref{SmallParticleApproximation}). We have generalized (\ref{BuoyancyChauchat}) to arbitrarily large particles (\S\ref{NewClosure}), resulting in a new closure given by either of the four equivalent expressions (\ref{BuoyancyNew1})-(\ref{BuoyancyNew3}). They involve the smoothing operator $\mathcal{L}$ defined in (\ref{L}) and simplified in (\ref{LA}) for systems of identically-sized spheres. It essentially represents a low-pass filter, attenuating buoyancy effects at short wavelengths or large wavenumber vectors $\bm{k}$ as $(R|\bm{k}|)^{-4}$. It is this attenuation that prevents buoyancy from causing Hadamard instabilities (\S\ref{Wellposedness}). We have also shown that the virtual-mass force density is suppressed in a similar manner and that it also does not introduce ill-posedness into two-fluid models.

In conclusion, the ill-posedness problem of two-fluid models \citep{Lhuillieretal13}, which was linked to the generalized-buoyancy and virtual-mass force densities, was actually never a problem to begin with. It was artificially created by employing mathematical closures for these densities that do not take into account the particle-size-related smoothing that inherently results from integrating macroscopic quantities such as $\bm{\nabla}\cdot\bm{\hat\sigma^\mathrm{f}}$ (buoyancy) and $\bm{u}_\mathrm{f}$ (virtual mass) over the particles' volumes. For example, when \citet{Lhuillieretal13} introduced the ill-posedness problem in their review, their very starting point, their equation~(2.2), was already inappropriate, and everything that followed was therefore unintentionally about how to treat the symptom rather than the cause.

From a purely practical viewpoint, when conducting two-fluid model simulations, it is often sufficient to employ the closure by \citet{RevilBaudardChauchat13}, (\ref{BuoyancyChauchat}). Only when the numerical mesh grid size is of the order of the particle size or smaller, it becomes necessary to account for the effects of the smoothing operator $\mathcal{L}$, which can be done in the manner described in \S\ref{TwoFluidNumericalImplementation}. By contrast, in the context of a theoretical linear instability analysis, and not only when looking for Hadamard instabilities, it is often crucial to consider the effects of $\mathcal{L}$, especially when small-wavelength perturbations are explored. And, lastly, an accurate expression for the generalized-buoyancy force density is also a prerequisite for empirically studying two-phase flow rheology with particle-resolving numerical simulations, especially for systems that lack scale separation, where the differences between (\ref{BuoyancyChauchat}) and our generalized closure can be considerable.

\backsection[Data and code availability]{Figures~\ref{ConcentrationVelocityProfiles}-\ref{Fig_BuoyancyClosures} can be produced from the MATLAB code and data made public by \citet{Pahtzetal26}. A MATLAB code producing figures~\ref{SigmaProfileszxSettling}-\ref{StatisticalBias} is available at \url{https://doi.org/10.5281/zenodo.21084179}.}

\backsection[Acknowledgment]{We acknowledge thought-provoking discussions with Nicolas Fintzi. We thank Jake Langham for discovering a serious mistake in an earlier version of the paper. We thank Michael P. Lamb for sharing their ideas and unpublished information regarding the paper by \citet{Lambetal17}. We thank Gary Parker and Giovanni Seminara for sharing their ideas and an unpublished manuscript on buoyancy.}

\backsection[Funding]{Z.H. acknowledges financial support from grant National Key R \& D Program of China (2023YFC3008100). R.Z. acknowledges financial support from grant National Natural Science Foundation of China (no.~12402494). T.P. acknowledges financial support from grants National Natural Science Foundation of China (nos.~12350710176, 12272344). Z.H. acknowledges financial support from grant National Natural Science Foundation of China (no.~12572470).}

\backsection[Declaration of interests]{The authors report no conflict of interest.}




\appendix
\section{Micromechanical expressions for the fluid and dispersed-phase stress tensors} \label{StressTensors}
The micromechanical expressions for the effective mechanical fluid phase and dispersed-phase stress tensors, respectively, are given by \citep[][and their Supplementary Material]{Pahtzetal26}
\begin{align}
 \bm{\sigma^\mathrm{f}}&\equiv\beta_\mathrm{f}\langle\bm{\sigma}\rangle^\mathrm{f}+\bm{\sigma^\mathrm{f}_\mathrm{s}}=\langle\bm{\sigma}\rangle^\mathrm{f}+\bm{\sigma^\mathrm{f}_\mathrm{s^\prime}}+\left[\langle\bm{\sigma}\rangle^\mathrm{f}\right]_\mathrm{s}^\prime, \label{SigmaFluid} \\
 \bm{\sigma^\mathrm{f}_\mathrm{s}}&\equiv\left\langle\sum\nolimits_pS^p\left(\underline{\bm{r}^p\bm{n}^p\cdot\bm{\sigma}\delta^p_{\mathrm{L}\bm{x}}}_{\mathbb{S}^p}-\underline{\bm{r}^p\delta^p_{\mathrm{L}\bm{x}}}_{\mathbb{V}^p}\underline{\bm{n}^p\cdot\bm{\sigma}}_{\mathbb{S}^p}\right)\right\rangle, \\
 \bm{\sigma^\mathrm{f}_\mathrm{s^\prime}}&\equiv\left\langle\sum\nolimits_pS^p\left(\underline{\bm{r}^p\bm{n}^p\cdot\left(\bm{\sigma}-\langle\bm{\sigma}\rangle^\mathrm{f}\right)\delta^p_{\mathrm{L}\bm{x}}}_{\mathbb{S}^p}-\underline{\bm{r}^p\delta^p_{\mathrm{L}\bm{x}}}_{\mathbb{V}^p}\underline{\bm{n}^p\cdot\left(\bm{\sigma}-\langle\bm{\sigma}\rangle^\mathrm{f}\right)}_{\mathbb{S}^p}\right)\right\rangle, \\
 \bm{\sigma^\mathrm{s}}&\equiv\left\langle\frac{1}{2}\sum\nolimits_{pq}\left(\bm{r}^{pq}_\mathrm{c}\delta^{pq}_{\mathrm{cL}\bm{x}}-\bm{r}^{qp}_\mathrm{c}\delta^{qp}_{\mathrm{cL}\bm{x}}\right)\bm{F}^{pq}_\mathrm{c}\right\rangle-\left\langle\sum\nolimits_p\underline{\bm{r}^p\delta^p_{\mathrm{L}\bm{x}}}_{\mathbb{V}^p}\bm{F}_\mathrm{c}^p\right\rangle.
\end{align}
These expressions involve the operators $\underline{\cdot}_{\mathbb{V}^p}\equiv\frac{1}{V^p}\int_{\mathbb{V}^p}\cdot\d V$ and $\underline{\cdot}_{\mathbb{S}^p}\equiv\frac{1}{S^p}\int_{\mathbb{S}^p}\cdot\d S$, with $V^p$ the volume and $S^p$ the surface area of a particle $p$, and the definitions
\begin{align}
 \delta^p_{\mathrm{L}\bm{x}}[\bm{y}]&\equiv\int_0^1\delta[\bm{x}-\bm{x}^p-\lambda(\bm{y}-\bm{x}^p)]\d\lambda, \\
 \delta^{pq}_{\mathrm{cL}\bm{x}}&\equiv\delta^p_{\mathrm{L}\bm{x}}\left[\bm{x}^{pq}_\mathrm{c}\right],
\end{align}
where $\delta$ denotes the multivariate delta distribution. For the volume and surface averages over terms involving $\delta^p_{\mathrm{L}\bm{x}}[\bm{y}]$, such as $\underline{\bm{r}^p\delta^p_{\mathrm{L}\bm{x}}}_{\mathbb{V}^p}$, $\bm{y}$, not $\bm{x}$, is the coordinate over which is averaged (i.e., the result of the averaging depends on $\bm{x}$). Furthermore, $\bm{\sigma^\mathrm{f}_\mathrm{s}}$ in (\ref{SigmaFluid}) has been decomposed into three contributions via $\bm{\sigma^\mathrm{f}_\mathrm{s}}=\beta_\mathrm{s}\langle\bm{\sigma}\rangle^\mathrm{f}+\bm{\sigma^\mathrm{f}_\mathrm{s^\prime}}+[\langle\bm{\sigma}\rangle^\mathrm{f}]_\mathrm{s}^\prime$, where $\bm{\sigma^\mathrm{f}_\mathrm{s^\prime}}$ has the important property that it vanishes when $\langle\bm{\sigma}\rangle^\mathrm{f}=\bm{\sigma}$ on all particles' surfaces, and where $[\cdot]_\mathrm{s}^\prime$ denotes an operator that acts on an arbitrary macroscopic tensor field $\bm{A}$ (i.e., fields that satisfy $\langle\bm{A}\rangle=\bm{A}$) in the following manner:
\begin{equation}
 [\bm{A}]_\mathrm{s}^\prime\equiv\left\langle\sum\nolimits_pV^p\underline{\bm{r}^p\left(\bm{\nabla}\cdot\bm{A}-\underline{\bm{\nabla}\cdot\bm{A}}_{\mathbb{V}^p}\right)\delta^p_{\mathrm{L}\bm{x}}}_{\mathbb{V}^p}\right\rangle. \label{[]}
\end{equation}
It satisfies (\ref{FluctuationOperator1}) and therefore implies
\begin{equation}
 \bm{\nabla}\cdot\left(\beta_\mathrm{s}\langle\bm{\sigma}\rangle^\mathrm{f}+\left[\langle\bm{\sigma}\rangle^\mathrm{f}\right]_\mathrm{s}^\prime\right)=\beta_\mathrm{s}\left\langle\sum\nolimits_p\frac{X^p}{V^p}\int_{\mathbb{S}^p}\bm{n}^p\cdot\langle\bm{\sigma}\rangle^\mathrm{f}\right\rangle^\mathrm{s}+\left\langle\bm{\sigma}^\mathrm{T}\right\rangle^\mathrm{f}\cdot\bm{\nabla}\beta_\mathrm{s}, \label{sigmafshelp}
\end{equation}
where $\cdot^\mathrm{T}$ denotes the transpose. Equation~(\ref{sigmafshelp}) leads to (\ref{sigmafs}) via (\ref{SigmaFluid}).

\section{Derivation of $\mathcal{L}$ for a mixture of continuous fluid and identically-sized spheres} \label{DerivationDifferentialOperator}
Here, we derive an expression for the operator $\mathcal{L}$ defined by (\ref{L}) for a system of identically-sized spherical particles of radius $R$ and volume $V=4\pi R^3/3$. To this end, we aim to manipulate $X^p$ and $\underline{\bm{A}}_{\mathbb{V}^p}$ appearing in (\ref{L}) in a manner that allows a straightforward application of the averaging rules in (\ref{AveragingProcedure}) and (\ref{ReynoldsRule}) afterwards. To manipulate $\underline{\bm{A}}_{\mathbb{V}^p}$, we start with a general expression for $\underline{(r^p_z)^k}_{\mathbb{V}^p}$, where $k\geq0$ is an integer and the $z$-direction is aligned with the polar axis in spherical coordinates:
\begin{equation}
 \underline{\left(r^p_z\right)^k}_{\mathbb{V}^p}=\frac{3}{4\pi R^3}\int_0^R\int_0^{2\pi}\int_0^\pi(r\cos\theta)^kr^2\sin\theta\d\theta\d\phi\d r=\frac{3\left(1+(-1)^k\right)R^k}{2(k+1)(k+3)}. \label{rpVp}
\end{equation}
Using (\ref{rpVp}) and exploiting spherical symmetry, one then obtains an expression for $\underline{\bm{A}}_{\mathbb{V}^p}$ from the Taylor expansion of $\bm{A}[\bm{x}]$ in $\bm{x}=\bm{x}^p$,
\begin{equation}
 \bm{A}[\bm{x}]=\sum_{k=0}^\infty\frac{1}{k!}(\bm{r}^p[\bm{x}])^{(k)}\odot\bm{\nabla}^{(k)}\bm{A}[\bm{x}^p]\Rightarrow\underline{\bm{A}}_{\mathbb{V}^p}=\sum_{k=0}^\infty\frac{3R^{2k}\Delta^k\bm{A}[\bm{x}^p]}{(2k+1)!(2k+3)},
\end{equation}
where $\cdot^{(k)}$ denotes the $k$-th power with respect to the dyadic product (e.g., $\bm{r}^{(3)}=\bm{r}\bm{r}\bm{r}$) and $\odot$ the full contraction operating, i.e., $\bm{A}^{(1)}\odot\bm{B}^{(1)}=\bm{A}\cdot\bm{B}$, $\bm{A}^{(2)}\odot\bm{B}^{(2)}=\bm{A}\bm{A}:\bm{B}\bm{B}$, and so forth. Defining the differential operator
\begin{equation}
 \mathcal{M}\equiv\sum_{k=0}^\infty a_kR^{2k}\Delta^k,\quad\text{with}\quad a_k\equiv\frac{3}{(2k+1)!(2k+3)}=\left(1,\frac{1}{10},\frac{1}{280},\frac{1}{15120},\dots\right), \label{M}
\end{equation}
we can therefore further express $\underline{\bm{A}}_{\mathbb{V}^p}$ as
\begin{equation}
 \underline{\bm{A}}_{\mathbb{V}^p}=\mathcal{M}\bm{A}[\bm{x}^p]. \label{AVp}
\end{equation}
The operator $\mathcal{M}$ also links the indicator function $X^p$ of a particle $p$ to the delta distribution $\delta[\bm{x}-\bm{x}^p]$ localized in $\bm{x}^p$:
\begin{equation}
 X^p=\mathcal{M}V\delta[\bm{x}-\bm{x}^p]. \label{XpDelta}
\end{equation}
This can be readily confirmed through application on a test function $\phi$ with compact support:
\begin{align}
 X^p[\phi]&=\int_{\mathbb{R}^3}X^p\phi\d^3x=\int_{\mathbb{V}^p}\phi\d V=V\underline{\phi}_{\mathbb{V}^p}, \\
 (\mathcal{M}V\delta[\bm{x}-\bm{x}^p])[\phi]&=(V\delta[\bm{x}-\bm{x}^p])[\mathcal{M}\phi]=V\mathcal{M}\phi[\bm{x}^p]=V\underline{\phi}_{\mathbb{V}^p}.
\end{align}
As a consequence of (\ref{AVp}) and (\ref{XpDelta}), the generalized function $X^p\underline{\bm{A}}_{\mathbb{V}^p}$ can be expressed as
\begin{equation}
 X^p\underline{\bm{A}}_{\mathbb{V}^p}=(\mathcal MV\delta[\bm{x}-\bm{x}^p])(\mathcal{M}\bm{A})[\bm{x}^p]=\mathcal MV\delta[\bm{x}-\bm{x}^p]\mathcal{M}\bm{A},
\end{equation}
which again is readily confirmed through application on a test function $\phi$:
\begin{equation}
\begin{split}
 &\left(\mathcal MV\delta[\bm{x}-\bm{x}^p](\mathcal{M}\bm{A})[\bm{x}^p]\right)[\phi]=\left(V\delta[\bm{x}-\bm{x}^p](\mathcal{M}\bm{A})[\bm{x}^p]\right)[\mathcal M\phi]\\
 &=\left(V\delta[\bm{x}-\bm{x}^p](\mathcal{M}\bm{A})[\bm{x}]\right)[\mathcal M\phi]=\left(\mathcal MV\delta[\bm{x}-\bm{x}^p]\mathcal{M}\bm{A}\right)[\phi].
\end{split}
\end{equation}
Hence, using the averaging rules in (\ref{AveragingProcedure}) and (\ref{ReynoldsRule}), we obtain
\begin{equation}
\begin{split}
 \mathcal{L}\beta_\mathrm{s}\bm{A}&=\left\langle\sum\nolimits_pX^p\underline{\bm{A}}_{\mathbb{V}^p}\right\rangle=\left\langle\sum\nolimits_p\mathcal MV\delta[\bm{x}-\bm{x}^p]\mathcal{M}\bm{A}\right\rangle \\
 &=\mathcal M\left\langle\sum\nolimits_pV\delta[\bm{x}-\bm{x}^p]\mathcal{M}\bm{A}\right\rangle=\mathcal M\left\langle\sum\nolimits_pV\delta[\bm{x}-\bm{x}^p]\right\rangle\mathcal{M}\bm{A} \\
 &=\mathcal{M}nV\mathcal{M}\bm{A}=\mathcal{M}\left(\mathcal{M}^{-1}\beta_\mathrm{s}\right)\mathcal{M}\bm{A}, \label{CalculationL}
\end{split}
\end{equation}
where we exploited that $\bm{A}$ is a macroscopic tensor field, $\mathcal{M}\bm{A}=\mathcal{M}\langle\bm{A}\rangle=\langle\mathcal{M}\bm{A}\rangle$, and introduced the number density $n$ defined as \citep{Pahtzetal26}
\begin{equation}
 n\equiv\left\langle\sum\nolimits_p\delta[\bm{x}-\bm{x}^p]\right\rangle=\mathcal{M}^{-1}\beta_\mathrm{s}/V. \label{NumberDensity}
\end{equation}
The equality in (\ref{NumberDensity}) follows from averaging (\ref{XpDelta}), where the inverse operator $\mathcal{M}^{-1}$ has the same structure as $\mathcal{M}$ in (\ref{M}),
\begin{equation}
 \mathcal{M}^{-1}=\sum_{k=0}^\infty b_kR^{2k}\Delta^k, \label{Minv}
\end{equation}
with coefficients $b_k$ that can be computed iteratively from $\sum_{i=0}^ka_ib_{k-i}=\delta_{k0}$ starting at $k=0$. Alternatively, from comparison with (\ref{Mhat}), $b_k$ are also the even Taylor coefficients of $1/M_{(-\mathrm{i}\alpha)}$ in $\alpha=0$:
\begin{equation}
 b_k=\frac{1}{(2k)!}\lim_{\alpha\to0}\frac{\d^{2k}}{\d\alpha^{2k}}\frac{\alpha^3}{3(\alpha\cosh\alpha-\sinh\alpha)}=\left(1,-\frac{1}{10},\frac{9}{1400},-\frac{19}{54000},\dots\right).
\end{equation}

\bibliographystyle{jfm}

\end{document}